\documentclass[12pt,epsf,amssymb]{article}
\usepackage{graphicx}

\def\lsim{\raise0.3ex\hbox{$<$\kern-0.75em\raise-1.1ex\hbox{$\sim$}}}
\def\gsim{\raise0.3ex\hbox{$>$\kern-0.75em\raise-1.1ex\hbox{$\sim$}}}
\def\noi{\noindent} \def\nn{\nonumber} \def\bea{\begin{eqnarray}}
\def\eea{\end{eqnarray}} \def\beq{\begin{equation}}
\def\eeq{\end{equation}} 
\def\beeq{\begin{eqnarray}} \def\eeeq{\end{eqnarray}} \def\R{ {\rm R
\kern -.31cm I \kern .15cm}} \def\C{ {\rm C \kern -.15cm \vrule
width.5pt \kern .12cm}} \def\Z{ {\rm Z \kern -.27cm \angle \kern
.02cm}} \def\N{ {\rm N \kern -.26cm \vrule width.4pt \kern .10cm}}
\def\1{{\rm 1\mskip-4.5mu l} }

\begin{document} \begin{center} 
{\large \bf Relation between Light Cone Distribution Amplitudes}\par \vskip 1 truemm

{\large \bf and Shape Function in B mesons} 

\vskip 1  truecm  
{\bf A. Le Yaouanc, L. Oliver and J.-C. Raynal}\par \vskip 3 truemm

Laboratoire de Physique Th\'eorique\footnote{Unit\'e Mixte de
Recherche UMR 8627 - CNRS }\\    Universit\'e de Paris XI,
B\^atiment 210, 91405 Orsay Cedex, France 
\end{center}

\begin{abstract}
The Bakamjian-Thomas relativistic quark model provides a Poincar\'e representation of bound states with a fixed number of constituents and, in the heavy quark limit, form factors of currents satisfy covariance and Isgur-Wise scaling. We compute the Light Cone Distribution Amplitudes of $B$ mesons $\varphi_{\pm}^B(\omega )$ as well as the Shape Function $S( \omega)$, that enters in the decay $B \to X_s \gamma$, that are also covariant in this class of models. The LCDA and the SF are related through the quark model wave function. The former satisfy, in the limit of vanishing constituent light quark mass, the integral relation given by QCD in the valence sector of Fock space. Using a gaussian wave function, the obtained $S(\omega )$ is identical to the so-called Roman Shape Function. From the parameters for the latter that fit the $B \to X_s\gamma$ spectrum we predict the behaviour of $\varphi_{\pm}^B(\omega)$. We discuss the important role played by the constituent light quark mass. In particular, although $\varphi_-^B(0) \not= 0$ for vanishing light quark mass, a non-vanishing mass implies the unfamiliar result $\varphi_-^B (0) = 0$. Moreover, we incorporate the short distance behaviour of QCD to $\varphi_+^B (\omega )$, which has sizeable effects at large $\omega$. We obtain the values for the parameters $\overline{\Lambda} \cong 0.35$~GeV and $\lambda_B^{-1} \cong 1.43$~GeV$^{-1}$.  We compare with other theoretical approaches and illustrate the great variety of models found in the literature for the functions $\varphi_{\pm}^B (\omega )$; hence the necessity of imposing further constraints as in the present paper. We briefly review also the different phenomena that are sensitive to the LCDA. The value that we find for $\lambda_B^{-1}$ fulfills the upper bound recently measured by BaBar.
\end{abstract}
%\vskip 0.5 truecm

\noi LPT Orsay 07-37  \par 
\noi June 2007\par 
\newpage
\pagestyle{plain} 
\section{Introduction.}
\hspace*{\parindent}
The Light Cone Distribution Amplitudes (LCDA) of heavy-light mesons $\Phi_+^B (\xi )$, $\Phi_-^B(\xi)$ \cite {1r} or, in the heavy quark limit $\varphi_+^B (\omega )$ and $\varphi_-^B (\omega )$ \cite{2r}, are fundamental functions that enter in the large energy limit of amplitudes of semileptonic decays \cite{2r} and in non-leptonic decays of $B$ mesons \cite{3r}, in the determination of the form factor $F_+^{B\to \pi}(0)$ \cite{4r} and, more directly, in the decay $B^- \to \gamma \ell \overline{\nu}_{\ell}$ \cite{5r,6r,7r,8r}.\par

On the theoretical side, these functions have been studied in the perturbative regime at large $\omega$ \cite{7r,9r}, and a number of very varied Ans\"atze have been proposed for the dominant part of them at low $\xi$ or $\omega$, where the function is peaked at $\xi \sim {\Lambda_{QCD} \over m_B}$ or $\omega \sim \Lambda_{QCD}$ \cite{2r,7r,8r}. \par

On the other hand, one can obtain model independent information on these functions from the measurement of the spectrum in the decay $B^- \to \gamma \ell \overline{\nu}_{\ell}$, that is directly related to one of the LCDA \cite{5r,6r,7r,8r}. \par

On the theoretical side, although rigorous results are known in the perturbative regime $\omega \gg \Lambda_{QCD}$, the guesses advanced for the main non-perturbative part of the LCDA amplitudes come essentially from QCD Sum Rules, imposing continuity between the perturbative and long distance regimes \cite{7r,9r}.\par

The motivation of the present work is to compute the LCDA in a class of relativistic quark models, namely the Bakamjian-Thomas (BT) quark models \cite{10r,11r,12r,13r,14r}. This is a class of models with a fixed number of constituents where the states are covariant under the Poincar\'e group. The model relies on an appropriate Lorentz boost of eigenfunctions of a Hamiltonian describing the spectrum at rest. On the other hand, one has demonstrated that the matrix elements of currents between hadrons are covariant in the heavy quark limit and exhibit Isgur-Wise scaling \cite{15r} in this limit \cite{14r}. Given a Hamiltonian describing the spectrum, the model provides an unambiguous result for the elastic Isgur-Wise function $\xi (w)$ \cite{14r,16r}. On the other hand, the sum rules (SR) in the heavy quark limit of QCD, like Bjorken or Uraltsev SR are satisfied in the model \cite{17r,18r}, as well as SR involving higher derivatives of $\xi (w)$ at zero recoil \cite{19r}.\par  

The interest of computing the LCDA functions $\varphi_{\pm}^B(\omega)$ in this framework is to directly relate them, in an unambiguous way, to the Shape Function $S(\omega )$ \cite{20r,21r,22r,23r,24r,25r,26r} in $B \to X_s\gamma$, that can also be computed in the BT class of models.\par

One could calculate the LCDA within the BT scheme using a Hamiltonian describing the spectrum, like the Godfrey-Isgur (GI) model \cite{27r}, that has been used elsewhere to compute the elastic Isgur-Wise function $\xi (w)$ and the inelastic ones $\tau_{1/2}(w)$, $\tau_{3/2}(w)$ \cite{16r}. However, we have tested the GI model to compute the Shape Function $S(\omega )$ and have realized that this model does not fit the $B \to X_s \gamma$ spectrum. This is the reason why we have decided another phenomenological approach, namely to relate the Light Cone Distribution Amplitudes $\varphi_{\pm}^B (\omega )$ to the Shape Function $S(\omega )$, a relation that is provided by the BT scheme. Using this relation, and a SF $S(\omega )$ that fits the $B \to X_s \gamma$ spectrum, one can predict the LCDA. The problems for the SF and also a discussion of heavy-to-light form factors within the BT scheme for the GI spectroscopic model will be given in detail elsewhere.\par

The paper is organized as follows. In Section~2 we give the master formulae defining the theoretical framework of BT quark models. In Section~3 we review the definitions of the LCDA at finite mass and in the heavy quark limit of QCD and in Section~4 we set a main natural hypothesis to compute the LCDA within quark models. In Section~5, to introduce the technicalities of the BT model, we review the calculation of the heavy meson decay constant. In Section~6 we obtain our main results, namely the expressions for the LCDA in the BT quark models. In Section~7 we compute the SF $S(\omega )$ in BT models and show that, in the case of the harmonic oscillator, it is identical with the so-called Roman Shape function, used to fit the $B \to X_s \gamma$ spectrum \cite{20r,21r,22r,26r}. In Section~8 we use the parameters of the latter to predict the LCDA functions $\varphi_{\pm}^B(\omega )$ and their moments. In Section~9, following Braun, Ivanov and Korchemsky \cite{7r} and Lee and Neubert \cite{9r}, we introduce the radiative tail of $\varphi_{+}^B(\omega )$. In Section~10 we compare our results with proposals for the LCDA in other theoretical schemes. In Section~11 we review the different phenomena that are sensitive to the LCDA, and in Section~12 we conclude.

\section{The Bakamjian-Thomas relativistic quark model.}
\hspace*{\parindent}
As explained in \cite{14r}, the construction of the BT wave function in motion involves a {\it unitary} transformation that relates the wave function $\Psi_{s_1, \cdots ,s_n}^{(P)}({\bf p}_1, \cdots , {\bf p}_n)$ in terms of one-particle variables, the spin ${\bf S}_i$ and momenta ${\bf p}_i$ to the so-called {\it internal} wave function $\Psi_{s_1, \cdots ,s_n}^{int}({\bf P}, {\bf k}_2, \cdots ,{\bf k}_n)$ given in terms of another set of variables, the total momentum ${\bf P}$ and the internal momenta ${\bf k}_1, {\bf k}_2, \cdots ,{\bf k}_n$ ($\sum\limits_i {\bf k}_i = 0$). This property ensures that, starting from an orthonormal set of internal wave functions, one gets an orthonormal set of wave functions in any frame. The base $\Psi_{s_1, \cdots , s_n}^{(P)}({\bf p}_1, \cdots , {\bf p}_n)$ is useful to compute one-particle matrix elements like current one-quark matrix elements, while the second  $\Psi_{s_1, \cdots ,s_n}^{int}({\bf P}, {\bf k}_2, \cdots ,{\bf k}_n)$ allows to exhibit Poincar\'e covariance. In order to satisfy the Poincar\'e commutators, the unique requirement is that the mass operator $M$, i.e. the Hamiltonian describing the spectrum at rest, should depend only on the internal variables and be rotational invariant, i.e. $M$ must commute with ${\bf P}$, ${\partial \over \partial {\bf P}}$ and ${\bf S}$. The internal wave function at rest $(2 \pi )^3 \delta ({\bf P}) \varphi_{s_1, \cdots , s_n}({\bf k}_2, \cdots , {\bf k}_n)$ is an eigenstate of $M$, ${\bf P}$ (with ${\bf P} = 0$), ${\bf S}^2$ and ${\bf S}_z$,  while the wave function in motion of momentum ${\bf P}$ is obtained by applying the boost ${\bf B}_P$, where $P^0 = \sqrt{{\bf P}^2 + M^2}$ involves the dynamical operator $M$. \par

The final output of the formalism that gives the total wave function in motion $\Psi_{s_1, \cdots ,s_n}^{(P)}({\bf p}_1, \cdots , {\bf p}_n)$ in terms of the internal wave function at rest $\varphi_{s_1, \cdots , s_n}({\bf k}_2, \cdots , {\bf k}_n)$ is the formula
\bea
\label{1e}
&&\Psi_{s_1, \cdots ,s_n}^{(P)}({\bf p}_1, \cdots , {\bf p}_n) = (2\pi )^3 \delta \left ( \sum_i {\bf p}_i - {\bf P}\right )  \sqrt{{\sum\limits_i p_i^0 \over M_0}}  \left ( \prod_i {\sqrt{k_i^0} \over \sqrt{p_i^0}}\right )\nn \\
&&\sum_{s'_1, \cdots , s'_n} \left [ D_i({\bf R}_i)\right ]_{s_i,s'_i} \varphi _{s'_1, \cdots , s'_n} ({\bf k}_2, \cdots , {\bf k}_n)
\eea

\noi where $p_i^0 = \sqrt{{\bf p}_i^2 + m_i^2}$ and $M_0$ is the free mass operator is given by $M_0 = \sqrt{(\sum\limits_i p_i)^2}$. \par

The internal momenta of the hadron at rest are given in terms of the momenta of the hadron in motion by the {\it free boost} $k_i = {\bf B}_{\sum\limits_i p_i}^{-1} p_i$ where the operator ${\bf B}_p$ is the boost $(\sqrt{p^2}, {\bf 0}) \to p$, the Wigner rotations ${\bf R}_i$ in the preceding expression ${\bf R}_i = {\bf B}_{p_i}^{-1}{\bf B}_{\sum\limits_i p_i}^{-1} {\bf B}_{k_i}$ and the states are normalized by $<{\bf P'}, {S'}_z| {\bf P}, {S}_z> \ = \ (2 \pi )^3 \delta ( {\bf P'} - {\bf P}) \delta_{S_z,S'_z}$.\par

The current one-quark matrix element acting on quark 1 between two hadrons is then given by the expression
\bea
\label{2e}
&&<\Psi ' ({\bf P'}, {S}'_z)| J^{(1)}| \Psi ({\bf P}, {S}_z)>\ = \int {d {\bf p'}_1 \over (2 \pi)^3}\ {d {\bf p}_1 \over (2 \pi)^3} \left ( \prod_{i=2}^n {d {\bf p}_i \over (2 \pi)^3}\right )\nn \\
&&\Psi_{s'_1, \cdots ,s_n}^{P'} ({\bf p'}_1 , \cdots , {\bf p}_n)^* < {\bf p'}_1, {s'}_1|J^{(1)}|{\bf p}_1, {s}_1>\Psi_{s_1, \cdots , s_n}^{P} ({\bf p}_1 , \cdots ,{\bf p}_n)
\eea

\noi where $\Psi_{s_1, \cdots , s_n}^{P} ({\bf p}_1 , \cdots , {\bf p}_n)$ is given in terms of the internal wave function by (\ref{1e}) and $< {\bf p'}_1, {\bf s'}_1|J^{(1)}|{\bf p}_1, {\bf s}_1>$ is the one-quark current matrix element.\par

As demonstrated in \cite{14r,28r}, in this formalism, in the heavy quark limit, current matrix elements are covariant and exhibit Isgur-Wise scaling, and one can compute Isgur-Wise functions like $\xi (w)$, $\tau_{1/2}(w)$, $\tau_{3/2}(w)$ \cite{16r}. \par

In the present paper, as far as the LCDA are concerned, we are dealing with current matrix elements between one meson and the vacuum, i.e. $<0|J|\Psi>$. In \cite{29r} such matrix elements were considered and it was demonstrated that the decay constants of heavy-light mesons are covariant -- independent of the frame -- in the heavy quark limit, and exhibit heavy quark scaling, i.e. $f_B \sqrt{m_B}$ is a constant in this limit. This quantity was calculated for various Hamiltonians describing the meson spectrum.\par

We want now to go beyond and compute the LCDA $\varphi_{\pm}^B(\omega )$ starting from the meson-to-vacuum matrix elements $<0|J|\Psi>$. We can here advance our main result, that is parallel to the one obtained for meson-to-meson current matrix elements. The LCDA are covariant in the heavy quark limit, and can therefore be computed without any arbitrary parameter once the Hamiltonian giving the internal wave function is given. The same statement holds for the Shape function $S(\omega )$ in $B \to X_s\gamma$, that can be expressed as a meson-to-meson matrix element $<\Psi |O|\Psi>$. For the latter we will use the harmonic oscillator that gives the Roman Shape Function \cite{20r,21r,22r,26r}. We will then use the phenomenological parameters of this function, that fit the $B \to X_s\gamma$ spectrum, to predict the LCDA.

\section{B meson Light Cone Distribution Amplitudes.}
\hspace*{\parindent}
Let us define the LCDA $\Phi_+^B(\xi )$, $\Phi_-^B(\xi)$ \cite{1r}
$$<0|\overline{q}(z)S_{n_-}(z,0)\Gamma b(0)|\overline{B}_d(P)>\Big |_{z_+=z_{\bot}}$$
\beq
\label{3e}
= - {f_B\over 4} \int_0^1 d \xi\ e^{-i\xi P_+z_-} Tr \left \{ \Gamma ({/\hskip-2.5 truemm P} + m_B ) \gamma_5 \left [ \Phi_+^B(\xi ) - {{/ \hskip-2.5 truemm n}_- \over v \cdot n_-}\ {\Phi_+^B(\xi ) - \Phi_-^B (\xi ) \over 2}\right ] \right \}
\eeq
\vskip 5 truemm

\noi where $S_{n_-}(z,0)$ is the Wilson line following the light-like four vector $n_-= (1,0,0,-1)$ ($n_-^2 = 0)$, $v$ is the $B$ four velocity $v = {P \over m_B}$, $\Gamma$ is an arbitrary Dirac matrix, and the center-of-mass motion is along the Oz axis. $P_+$, $z_-$ are light-cone variables defined for any four vector $p$ by $p_+ = {p^0+ p^z \over \sqrt{2}}$, $p_- = {p^0 - p^z \over \sqrt{2}}$. Sometimes one takes $v\cdot n_- = 1$, i.e. the $B$ rest frame, but to exhibit covariance we have adopted a general value for $v\cdot n_-$. The LCDA $\Phi_{B_1}(\xi )$, $\Phi_{B_2}(\xi )$ satisfy the normalization conditions
\beq
\label{4e}
\int_0^1 d\xi \ \Phi_+^B(\xi ) = \int_0^1 d\xi \ \Phi_-^B(\xi ) = 1
\eeq

In the heavy quark limit $m_b \to \infty$ it is useful to use a new variable
\beq
\label{5e}
\omega = m_b \xi
\eeq

\noi keeping $\omega$ fixed, that yields the definition of the LCDA \cite{2r,4r,30r},
\beq
\label{6e}
<0|\overline{q}(z)S_{n_-}(z,0)\Gamma h_v(0)|\overline{B}_d(v)>\Big |_{z_+=z_{\bot}}
\eeq
$$= - {f_Bm_B \over 4} \int_0^{\infty} d \omega\ e^{-i\omega v_+z_-} Tr \left \{ \Gamma (1 + {/\hskip-2.5 truemm v} ) \gamma_5 \left [ \varphi_+^B(\omega) - {\varphi_+^B(\omega)  - \varphi_-^B(\omega) \over 2}\  {{/ \hskip-2.5 truemm n}_- \over v \cdot n_-}\right ] \right \}$$

\noi The relation between $\Phi_\pm ^B(\xi)$ and $\varphi_\pm ^B(\omega)$ is
\beq
\label{7e}
\varphi_\pm ^B(\omega) = {1 \over m_B} \Phi_\pm ^B \left ( {\omega \over m_B}\right )
\eeq

\noi and $\varphi_\pm ^B(\omega)$ satisfy the normalization conditions
\beq
\label{8e}
\int_0^{\infty} d \omega \ \varphi_\pm ^B(\omega) = 1
\eeq

\section{LCDA in quark models.}
\hspace*{\parindent}
In what follows, we will use the preceding relations to obtain the expression of $\varphi_\pm ^B(\omega)$ in BT quark models. In the class of BT quark models, gluon exchange is included in the potential. Therefore, in a way, the gluon field is integrated out, but one looses the explicit gauge invariance that is ensured by the Wilson line of the preceding expressions.\par

In a quark model, what we can consider is the matrix element involving constituent quarks, in particular constituent light quarks with a dynamical mass. Our ansatz will be to identify the QCD matrix element with the Wilson line with a matrix element involving the constituent quark field, or in more rigorous terms, one would say that one works in the light-cone gauge,
\beq
\label{9e}
A_+ = 0 \qquad\qquad\qquad S_{n_-}(z,0) = 1
\eeq

\noi and set
\bea
\label{10e}
&&<0|\overline{q}(z)S_{n_-}(z,0)\Gamma b(0)|\overline{B}_d(v)>\Big |_{z_+=z_{\bot}=0}\nn \\
&&= \ <0|\overline{q}_{constituent}(z)\Gamma b(0)|\overline{B}_d(v)>\Big |_{z_+=z_{\bot}=0}
\eea

\noi This is our main hypothesis and the starting point of the quark model calculation, that then follows in a straightforward way. From now on the constituent light quark field $q_{constituent}$ will be denoted by $q$. Of course, the condition (\ref{9e}) can hold only in field theory, and our BT scheme is just a model. Defining $\Phi_{\pm}(\xi )$ by the quark model expression~: 
$$<0|\overline{q}(z)\Gamma b(0)|\overline{B}_d(P)>\Big |_{z_+=z_{\bot}}$$
\beq
\label{11e}
= - {f_B\over 4} \int_0^1 d \xi\ e^{-i\xi P_+z_-} Tr \left \{ \Gamma ({/\hskip-2.5 truemm P} + m_B ) \gamma_5 \left [ \Phi_+^B(\xi ) - {{/ \hskip-2.5 truemm n}_- \over v \cdot n_-}\ {\Phi_+^B(\xi ) - \Phi_-^B (\xi ) \over 2}\right ] \right \}
\eeq
\vskip 5 truemm

\noi where $n_- = (1,0,0,-1)$, one obtains
\beq
\label{12e}
\Phi_+ ^B(\xi) = {1 \over f_Bm_B} \ {P_+ \over 2 \pi} \int dz_-\ e^{i\xi P_+z_-} <0|\overline{q}(z){{/ \hskip-2.5 truemm n}_- \over v \cdot n_-}\gamma_5 b(0)|\overline{B}_d(P)>\Big |_{z_+=z_{\bot}=0}
\eeq
$$\Phi_- ^B(\xi) = {1 \over f_Bm_B} \ {P_+ \over 2 \pi} \int dz_-\ e^{i\xi P_+z_-} <0|\overline{q}(z)\left ( 2 {/ \hskip-2.5 truemm v}- {{/ \hskip-2.5 truemm n}_- \over v \cdot n_-}\right ) \gamma_5 b(0)|\overline{B}_d(P)>\Big |_{z_+=z_{\bot}=0}$$
\vskip 5 truemm

Calling $p_2$ the four- momentum of the light quark, using translational invariance and integrating over $z_-$, these expressions write
\bea
\label{13e}
&&\Phi_+ ^B(\xi) = {1 \over f_Bm_B} \ <0|\overline{q}(0)\delta \left ( \xi - {p_{2+} \over P_+}\right ) {{/ \hskip-2.5 truemm n}_- \over v \cdot n_-}\gamma_5 b(0)|\overline{B}_d(P)>\nn \\
&&\Phi_- ^B(\xi) = {1 \over f_Bm_B} \ <0|\overline{q}(0)\delta \left ( \xi - {p_{2+} \over P_+}\right ) )\left ( 2 {/ \hskip-2.5 truemm v}-{{/ \hskip-2.5 truemm n}_- \over v \cdot n_-}\right ) \gamma_5 b(0)|\overline{B}_d(P)>
\eea

These will be the starting formulas to compute these functions in the BT quark model, from which we will deduce their heavy quark limit $\varphi_{\pm}^B(\omega )$. But let us first compute the heavy meson decay constant $f_B$ in the BT quark models, that will provide the desired normalization for the LCDA.

\section{B decay constant in BT models.}
\hspace*{\parindent}
The calculation of the matrix elements to obtain the LCDA $\varphi_{\pm}^B(\omega )$ is just reminiscent of the one made to obtain the corresponding decay constant \cite{29r}. In this latter case one needs the matrix element
\bea
\label{14e}
&&<0|\overline{q}(0)\Gamma b(0)|\overline{B}_d(v)>\nn \\
&&= \sqrt{N_c} \int {d{\bf p}_2 \over ( 2 \pi )^3} \sqrt{{\sum_i p_i^0 \over M_0}} \sqrt{{k_1^0 k_2^0 \over p_1^0p_2^0}} \sqrt{{m_1m_2 \over p_1^0 p_2^0}}\ {1 \over \sqrt{2}} \ \varphi ({\bf k}_2) \nn \\
&&Tr \left [ \Gamma \gamma^5 {\bf B}_u {\bf B}_{k_2} {1 + \gamma^0 \over 2} {\bf B}_u^{-1} {\bf B}_u {1 + \gamma^0 \over 2} {\bf B}_{k_1}^{-1} {\bf B}_u^{-1}\right ] 
\eea

\noi where $\varphi ({\bf k}_2)$ is the internal wave function at rest, with the normalization
\beq
\label{15e}
\int {d{\bf k}_2 \over (2 \pi )^3}\ | \varphi ({\bf k}_2)|^2 = 1
\eeq

\noi $p_1$ and $p_2$ ($m_1$ and $m_2$) are the quark four-momenta (masses ) of respectively the heavy and light quarks, ${\bf B}_p$ is the $4 \times 4$ boost matrix associated with the four-vector $p$, and the four vector $u$, $M_0$ and the relation between $k_i$ and $p_i$ are given by
\beq
\label{16e}
u = {p_1 + p_2 \over M_0} \qquad\qquad M_0 = \sqrt{(p_1 + p_2)^2}\qquad\qquad  {\bf B}_uk_i = p_i \ (i = 1, 2)
\eeq

\noi where $k_1$ and $k_2$ are the four-momenta of the quarks in the rest frame of the $B$ meson and ${\bf B}_u$ is the boost associated with the four-vector $u$. The products of $4 \times 4$ matrices under the trace read
\bea
\label{17e}
&&{\bf B}_u {\bf B}_{k_2} {1 + \gamma^0 \over 2} {\bf B}_u^{-1} = { m_2 + {/ \hskip-2.5 truemm p}_2 \over \sqrt{2m_2 \left  ( k_2^0 + m_2 \right ) }} {1 + {/ \hskip-2.5 truemm u} \over 2} \nn \\
&&{\bf B}_u  {1 + \gamma^0 \over 2} {\bf B}_{k_1}{\bf B}_u^{-1} = {1 + {/ \hskip-2.5 truemm u} \over 2} \ { m_1 + {/ \hskip-2.5 truemm p}_1 \over \sqrt{2m_1 \left ( k_1^0 + m_1 \right )}}
\eea

\noi This yields the expression
\bea
\label{18e}
&&<0|\overline{q}(0)\Gamma b(0)|\overline{B}_d(P)>\nn \\
&&= - {\sqrt{N_c}\over 8}  \int {d{\bf p}_2 \over ( 2 \pi )^3}\  \sqrt{2} \ {\sqrt{u^0} \over  p_1^0 p_2^0} \sqrt{{k_1^0 k_2^0 \over \left ( k_1^0 + m_1\right ) 
\left ( k_2^0 + m_2\right )}}\nn \\ 
&&Tr \left [  \gamma^5\Gamma \left ( m_2 + {/ \hskip-2.5 truemm p}_2 \right ) (1 + {/ \hskip-2.5 truemm u}) \left ( m_1 + {/ \hskip-2.5 truemm p}_1 \right )\right ]\varphi ({\bf k}_2)
\eea

\noi Using the current $\Gamma = \gamma_{\mu}\gamma_5$ we obtain, after some algebra and the definition of the four vector $u$ (\ref{16e}), 
\bea
\label{19e}
&&<0|\overline{q}(0)\gamma_{\mu}\gamma_5  b(0)|\overline{B}_d(P)> = {\sqrt{N_c}\over \sqrt{2}} \int {d{\bf p}_2 \over ( 2 \pi )^3} {\sqrt{u^0} \over  p_1^0 p_2^0} \sqrt{{k_1^0 k_2^0 \over \left ( k_1^0 + m_1\right ) 
\left ( k_2^0 + m_2\right )}} \nn \\
&&\left ( 1 + {m_1 + m_2 \over M_0}\right ) \left ( p_{1\mu} m_2 + p_{2\mu} m_1 \right ) \varphi ({\bf k}_2)
\eea

\noi Since the BT states are normalized acccording to
\beq
\label{20e}
< \overline{B}_d(P')| \overline{B}_d(P)>_{BT} \ = (2\pi)^3 \delta ({\bf P} - {\bf P'})
\eeq

\noi while the covariant normalization is $< \overline{B}_d(P')| \overline{B}_d(P)> = (2\pi)^3 2P^0 \delta ({\bf P} - {\bf P'})$, we have to identify the former matrix element with the definition of the decay constant
\beq
\label{21e}
<0|\overline{q}(0)\gamma_{\mu}\gamma_5  b(0)|\overline{B}_d(P)> = {f_B P_{\mu} \over \sqrt{2P^0}}
\eeq

\noi one obtains
\bea
\label{22e}
&&f_B \sqrt{m_B}  = \sqrt{N_c} \sqrt{v^0} \int {d{\bf p}_2 \over ( 2 \pi )^3} {\sqrt{u^0} \over  p_1^0 p_2^0} \sqrt{{k_1^0 k_2^0 \over \left ( k_1^0 + m_1\right ) 
\left ( k_2^0 + m_2\right )}} \left ( 1 + {m_1 + m_2 \over M_0}\right ) \nn \\
&&\left [ m_1 \left ( p_2 \cdot v\right ) + m_2 \left ( p_1 \cdot v\right )  \right ] \varphi ({\bf k}_2)
\eea

This expression is not covariant, but becomes covariant in the heavy quark limit. For $m_1 \to \infty$ one has
\beq
\label{23e}
m_0 \to m_1 \to m_B \ , \quad u \to v\ , \quad {p_1 \over m_1} \to v\ , \quad {k_1^0 \over m_1} \to 1\ , \quad k_2^0 \to p_2\cdot v
\eeq

\noi where $v$ is the $B$ meson four-velocity, and one gets the expression of the $B$ decay constant in the BT model in the heavy quark limit \cite{29r}, 
\beq
\label{24e}
f_B \sqrt{m_B}  = \sqrt{2} \sqrt{N_c}  \int {d{\bf p}_2 \over ( 2 \pi )^3} {1 \over p_2^0} \sqrt{\left ( p_2 \cdot v\right ) \left ( p_2 \cdot v+ m_2 \right )}\  \varphi \left ( \sqrt{\left ( p_2 \cdot v\right )^2 - m_2^2 }\right )
 \eeq

\noi This expression is covariant, satisfies heavy quark scaling and gives the decay constant in terms of the internal wave function. In the $B$ rest frame one gets
\beq
\label{25e}
f_B \sqrt{m_B}  =  {\sqrt{N_c}  \over \sqrt{2}} \ {1 \over \pi^2} \int_0^{\infty} dp \ {p}^2 \left ( {m_2 +  \sqrt{p^2 + m_2^2} \over  \sqrt{p^2 + m_2^2}}\right )^{1/2} 
\varphi (p)
 \eeq
 
 We have checked that these expressions for the decay constant hold exactly in the equivalent light-front approach of Cardarelli at al. \cite{13r} in the heavy mass limit.

\section{LCDA in the heavy quark limit in BT models.}
\hspace*{\parindent}
According to (\ref{13e}), we have to compute a generic matrix element
\beq
\label{26e}
\Phi_B(\xi) = {1 \over f_Bm_B}  <0|\overline{q}(0)\delta \left ( \xi - {p_{2+} \over P_+}\right )\Gamma  b(0)|\overline{B}_d(P)>\Big |_{z_+=z_{\bot}=0}
\eeq

\noi where $\Gamma$ is a Dirac matrix. In the BT quark model, this expression writes, taking into account the appropriate normalizations,
\bea
\label{27e}
&&\Phi(\xi) = - \sqrt{2P^0} {1 \over f_Bm_B} {\sqrt{N_c} \over 8} \int {d{\bf p}_2 \over ( 2 \pi )^3} {1 \over p_2^0} \delta \left ( \xi - {p_{2+} \over P_+}\right )  \\
&&= \sqrt{2} {\sqrt{u^0} \over  p_1^0} \sqrt{{k_1^0 k_2^0 \over \left ( k_1^0 + m_1\right ) 
\left ( k_2^0 + m_2\right )}} Tr \left [  \gamma^5\Gamma \left ( m_2 + {/ \hskip-2.5 truemm p}_2 \right ) (1 + {/ \hskip-2.5 truemm u})\left ( m_1 + {/ \hskip-2.5 truemm p}_1 \right )\right ]\varphi ({\bf k}_2)\nn 
\eea

Changing the measure and integrating relatively to $p_{2+}$ and $p_{2-}$, one obtains
$$\Phi(\xi) = - \sqrt{2P^0} {1 \over f_Bm_B} {\sqrt{N_c} \over 8} {1 \over \xi} {1 \over 2 \pi} \Big [ \int {d{\bf p}_{2\bot} \over ( 2 \pi )^2} \sqrt{2} {\sqrt{u^0} \over  p_1^0} \sqrt{{k_1^0 k_2^0 \over \left ( k_1^0 + m_1\right ) 
\left ( k_2^0 + m_2\right )}}$$
\beq
\label{28e}
Tr \left [  \gamma^5\Gamma \left ( m_2 + {/ \hskip-2.5 truemm p}_2 \right ) (1 + {/ \hskip-2.5 truemm u}) \left ( m_1 + {/ \hskip-2.5 truemm p}_1 \right )\right ]\varphi ({\bf k}_2)\Big ]_{p_{2+} = \xi  P_+,p_{2-} = {({\bf p}_{2\bot})^2 + m_2^2  \over 2P_+ \xi}} 
\eeq

\noi Making use of the definition of the four-vector $u$ (\ref{16e}) and computing the trace particularizing respectively to $\Gamma = {{/ \hskip-2 truemm n}_- \over v \cdot n_-}\ \gamma_5$ and $\Gamma = \left ( 2 {/ \hskip-2.5 truemm v}-{{/ \hskip-2 truemm n}_- \over v \cdot n_-}\right )\gamma_5$, one gets 
\bea
\label{29e}
&&\Phi^B_+(\xi) =  \sqrt{2P^0} {1 \over f_Bm_B} {\sqrt{N_c} \over 2} {1 \over \xi} {1 \over 2 \pi} \left \{ \int {d{\bf p}_{2\bot} \over ( 2 \pi )^2} \sqrt{2} {\sqrt{u^0} \over  p_1^0} \sqrt{{k_1^0 k_2^0 \over \left ( k_1^0 + m_1\right ) 
\left ( k_2^0 + m_2\right )}}\right .\nn \\
&&\left . \left ( 1 + {m_1 + m_2 \over M_0}\right ) \left [ {m_1 (p_2 \cdot n_-) + m_2 (p_1 \cdot n_-) \over v \cdot n_-}\right ] \varphi ({\bf k}_2)\right ]_{p_{2+} = \xi  P_+,p_{2-} = {({\bf p}_{2\bot})^2 + m_2^2  \over 2P_+ \xi}}\nn \\
&&\Phi^B_-(\xi) = \sqrt{2P^0} {1 \over f_Bm_B} {\sqrt{N_c} \over 2} {1 \over \xi} {1 \over 2 \pi} \left [ \int {d{\bf p}_{2\bot} \over ( 2 \pi )^2} \sqrt{2} {\sqrt{u^0} \over  p_1^0}\right . \nn \\
&&\sqrt{{k_1^0 k_2^0 \over \left ( k_1^0 + m_1\right ) \left ( k_2^0 + m_2\right )}} \left ( 1 + {m_1 + m_2 \over M_0}\right ) \left \{ 2\left [ m_1 (p_2 \cdot v) + m_2 (p_1 \cdot v)\right  ] \right .\nn \\
&&- \left . \left .   {m_1 (p_2 \cdot n_-) + m_2 (p_1 \cdot n_-) \over v \cdot n_-} \right \}  \varphi ({\bf k}_2)\right ]_{p_{2+} = \xi  P_+,p_{2-} = {({\bf p}_{2\bot})^2 + m_2^2  \over 2P_+ \xi}}
\eea

\subsection{Heavy quark limit.}
\hspace*{\parindent}
At finite mass the expressions (\ref{29e}) are not covariant. In the heavy quark limit, using now the variable $\omega = m_1 \xi$,
\bea
\label{30e}
&u \to v_1 \to v \qquad &\qquad\qquad  k_1^0 \to m_1\ , \ k_2^0 \to p_2 \cdot v \nn \\
&M_0 \to m_1 \to m_B &\qquad\qquad  m_1 \xi \to \omega 
\eea

\noi one obtains, denoting ${\bf p}_{2\bot} = {\bf p}_{\bot}$
\bea
\label{31e}
&&\varphi_+^B(\omega ) = {\sqrt{N_c} \over f_B\sqrt{m_B}} \sqrt{2}  {1 \over 8 \pi^2} \int d{\bf p}^2_{\bot}\nn \\
&&\sqrt{{{\bf p}^2_{\bot} + m_2^2 + \omega^2 \over {\bf p}^2_{\bot} + (\omega + m_2)^2 }} {\omega + m_2 \over \omega} \varphi \left ( {\sqrt{\left ( {\bf p}^2_{\bot} + \omega^2 + m_2^2 \right )^2 - 4 \omega^2 m_2^2} \over 2 \omega}\right )\nn \\
&&\varphi_-^B(\omega ) = {\sqrt{N_c} \over f_B\sqrt{m_B}} \sqrt{2}  {1 \over 8 \pi^2} \int d{\bf p}^2_{\bot} \sqrt{{{\bf p}^2_{\bot} + \omega^2 + m_2^2 \over {\bf p}^2_{\bot} + (\omega + m_2)^2 }} {{\bf p}^2_{\bot} +m_2(\omega + m_2) \over \omega^2}\nn \\
&&\varphi \left ( {\sqrt{\left ( {\bf p}^2_{\bot} + \omega^2 + m_2^2 \right )^2 - 4 \omega^2 m_2^2} \over 2 \omega}\right )
\eea 

\noi In the heavy quark limit, the LCDA $\varphi_{\pm}^B (\omega )$ are covariant since the boost is along Oz, and the variable $\omega$ can be written in the covariant form, 
\beq
\label{32e}
\omega = m_B \xi = m_B {p_{2+} \over P_+} = m_B {p_2 \cdot n_- \over P \cdot n_-} = {p_2 \cdot v_- \over v \cdot n_-}
\eeq

\noi Performing the change of variables
\beq
\label{33e}
p = {\sqrt{\left ( {\bf p}^2_{\bot} + \omega^2 + m_2^2 \right )^2 - 4 \omega^2 m_2^2} \over 2 \omega}
\eeq

\noi one obtains
\bea
\label{34e}
&&\varphi_+^B(\omega ) = {\sqrt{N_c} \over f_B\sqrt{m_B}} \sqrt{2}  {1 \over 4 \pi^2} (\omega + m_2) \int_{p_0 (\omega )}^{\infty} dp {p \over (p^2 + m_2^2)^{1/4} \left ( \sqrt{p^2 + m_2^2} + m_2\right )^{1/2}} \varphi (p)\nn \\
&&\varphi_-^B(\omega ) = {\sqrt{N_c} \over f_B\sqrt{m_B}} \sqrt{2}  {1 \over 4 \pi^2}  \int_{p_0 (\omega )}^{\infty} dp {p \left ( 2 \sqrt{p^2 + m_2^2} - \omega + m_2\right )  \over (p^2 + m_2^2)^{1/4} \left ( \sqrt{p^2 + m_2^2} + m_2\right )^{1/2}} \varphi (p)
\eea 

\noi with
\beq
\label{35e}
p_0 (\omega ) = {|\omega - m_2 | (\omega + m_2) \over 2 \omega}
\eeq

\noi One can check the normalization (\ref{8e}) by changing the order of the integrals
\beq
\label{36e}
\omega > 0, \quad p > p_0 (\omega )\qquad  \Leftrightarrow \qquad p> 0 , \quad \omega_- (p) < \omega < \omega_+ (p)
\eeq

\noi with
\beq
\label{37e}
\omega_{\pm} (p) = \sqrt{p^2 + m_2^2} \pm p
\eeq

\noi The integrals over $\omega$ are trivial
\beq
\label{38e}
\int_{\omega_-(p)}^{\omega_+(p)} d \omega (m_2 + \omega ) = \int_{\omega_-(p)}^{\omega_+(p)} d \omega \left ( 2 \sqrt{p^2 + m_2^2} - \omega + m_2 \right ) = 2p \left ( \sqrt{p^2 + m_2^2} + m_2\right )  
\eeq

\noi and the normalization (\ref{8e}) follows.

\subsection{Limit of vanishing light quark mass.}
\hspace*{\parindent}
In the limit of vanishing light quark mass, one can immediately demonstrate from expressions (\ref{34e}) that $\varphi_{\pm}^B (\omega )$ satisfy the differential equation, the so-called Wandzura-Wilczek approximation \cite{4r} \cite{31r}
\beq
\label{39e}
{d \varphi_-^B (\omega ) \over d \omega} = - {\varphi_+^B (\omega ) \over \omega}
\eeq

\noi or, equivalently, the integral relation
\beq
\label{40e}
\varphi_-^B (\omega ) = \int_{\omega}^{\infty} d\rho\ {\varphi_+^B (\rho ) \over \rho}
\eeq

\noi Moreover, for vanishing light quark mass, one gets from (\ref{34e}), for small $\omega$, 
\beq
\label{41e}
\varphi_+^B (\omega ) \sim \omega\qquad \varphi_-^B (\omega )\sim \hbox{Constant}\qquad (m_2 = 0)
\eeq

The relation (\ref{39e}) or (\ref{40e}) holds in QCD if one restricts to the valence quark sector of Fock space \cite{4r,31r}. It is reassuring that this relation holds also in the BT quark model, since it is obviously formulated in the valence quark approximation. For this relation to be satisfied one needs nevertheless the dynamical light quark mass to vanish. However, it will be interesting to grasp the significance of the corrections due to a non-vanishing $m_2$ and check its numerical effect on $\varphi_\pm^B (\omega )$, as we will do below.

\subsection{Behaviour of $\varphi_\pm^B (\omega )$ for non-vanishing light quark mass.}
\hspace*{\parindent}
A specific trend of our results for the LCDA $\varphi_\pm^B (\omega )$ is the important role played by the light quark mass $m_2$, as will show the numerical results of Section~8. The light quark mass has dramatic consequences, namely that the first derivative of $\varphi_+^B (\omega )$ and $\varphi_-^B (\omega )$ vanish, as we can see by inspection of formulas (\ref{34e}), (\ref{35e})~:  
\beq
\label{42e}
\varphi_+^{B'} (0 ) = 0 \qquad\qquad \varphi_-^B (0) = 0
\eeq

\noi This is apparently at odds with the general belief, based on QCDSR, that predicts the behaviour (\ref{41e}) \cite{2r,6r,7r,8r}. The relation (\ref{40e}) is strongly violated in the presence of a constituent light quark mass, since the l.h.s. of (\ref{40e}) vanishes~: $\varphi_-^B(\omega )$ vanishes for $\omega \to 0$, since the lower limit (\ref{35e}) of the integral behaves like 
\beq
\label{43NEWEQ}
p_0 (\omega ) \sim {m_2^2 \over \omega } \to \infty \qquad \qquad \hbox{for \quad $\omega \to 0$}
\eeq

In the QCDSR approach one has an expression giving the correction to the behaviour (\ref{41e}) from the $<\overline{q}q>$ condensate, i.e. the first order correction due to the {\it dynamical} light quark mass, the gap induced by dynamical chiral symmetry breaking. This is the constituent mass involved in our quark model calculation, proportional to the condensate. The corrections at first order in $<\overline{q}q>$ have been given by Grozin and Neubert \cite{2r} that write, in a simplified notation,
\bea
\label{43e}
&&\varphi_+^B (\omega ) = \varphi_+^{B(0)} (\omega )\    -  \   <\overline{q}q> \widetilde{f}(\omega )\nn \\
&&\varphi_-^B (\omega ) = \varphi_-^{B(0)} (\omega )\  -  \   <\overline{q}q> \widetilde{f}(\omega )
\eea

\noi  where $\varphi_\pm^B (\omega )$ are LCDA at leading order, independent of $<\overline{q}q>$, satisfying (\ref{39e}), and the function $\widetilde{f}(\omega )$ has the behaviour
\beq
\label{44e}
\widetilde{f}(0) = \widetilde{f}(\infty ) = 0
\eeq

\noi Notice moreover that the correction dependent of $<\overline{q}q>$ is the same for $\varphi_+^B (\omega )$ and $\varphi_-^B (\omega )$.\par

In our scheme, the first order corrections in $m_2$ to $\varphi_+(\omega )$ and $\varphi_- (\omega )$ are also equal, but our calculations contain higher orders in the constituent light quark mass, as we observe from (\ref{31e}), (\ref{34e}), (\ref{35e}), giving the much stronger behaviour (\ref{42e}). 

\subsection{Moments of $\varphi_{\pm}^B(\omega )$.}
\hspace*{\parindent}
Defining the moments
\beq
\label{47e}
M_+^{(N)} = \int_0^{\infty} d\omega \ \omega^N \varphi_+^B (\omega ) \qquad\qquad M_-^{(N)} = \int_0^{\infty} d\omega \ \omega^N \varphi_- ^B(\omega )
\eeq

\noi one finds for $m_2 \not= 0$ that $M_+^{(0)} = M_-^{(0)} = 1$, i.e. the normalization condition (\ref{8e}). Moreover, one finds for any moment with $N \geq 0$, for vanishing light quark mass $m_2 = 0$,
\beq
\label{48e}
M_+^{(N)} = (N+1) M_-^{(N)}
\eeq

\noi that holds obviously in the BT class of quark models since it follows from the QCD relation (\ref{39e}). Of interest are the moments $M_+^{(1)}$, $M_-^{(1)}$ that are given in the valence sector of QCD by \cite{2r}
\beq
\label{49e}
M_+^{(1)} = 2M_-^{(1)} = {4 \overline{\Lambda} \over 3}
\eeq

\noi and should allow to compute $\overline{\Lambda}$ in the BT models.\par

The moment $M_+^{(-1)} = \lambda_B^{-1}$, satisfies, due to the positivity condition in the absence of a radiative tail $\varphi_+(\omega ) \geq 0$ \cite{5r},
\beq
\label{49ebis}
\lambda_B^{-1} = M_+^{(-1)} \geq {1 \over M_+^{(1)} } = {3 \over 4\overline{\Lambda}} 
\eeq

\section{The Shape Function S($\omega$) in the BT approach.}
\subsection{Definitions of the Shape Function.}
\hspace*{\parindent}
The shape function $S(\omega )$ that enters in the description of the decay $B \to X_s\gamma$ \cite{22r,23r} is defined by the expression
\beq
\label{50e}
S(\omega ) \ {1 \over 2} \ Tr \left ( \Gamma {1 + {/ \hskip -2 truemm v} \over 2}\right ) = {<\overline{B}(p_B) | \overline{h}_v\Gamma \delta (\omega - i D_+) h_v | \overline{B}(p_B)> \over 2m_B}
\eeq

\noi where $p_+ = p^0 + p^z$ and the support of $S(\omega )$ is
\beq
\label{51e}
- \infty < \omega \leq \overline{\Lambda}
\eeq

\noi One uses also another definition \cite{24r,25r}
\beq
\label{52e}
\widehat{S}(\widehat{\omega}) = S (\overline{\Lambda} - \widehat{\omega})
\eeq

\noi with the support 
\beq
\label{53e}
0 \leq \widehat{\omega} < \infty
\eeq

The functions $S(\omega)$ or $\widehat{S}(\widehat{\omega})$ can also be computed in the BT model.

\subsection{Calculation of the Shape Function in the BT model.}
\hspace*{\parindent}
Since the field $h_v$ annihilates the $b$ quark within the $\overline{B}$ meson, one can introduce a complete set of intermediate states of the spectator quark $|p_2,s_2><p_2,s_2|$ in the preceding expression
\bea
\label{54e}
&&<\overline{B}(p_B) | \overline{h}_v(x)  \delta (\omega - i D_+) h_v(x) | \overline{B}(p_B)> \ = \nn \\
&&\sum_{p_2,s_2} \  <\overline{B}(p_B) | \overline{h}_v(x)|p_2,s_2><p_2,s_2|  \delta (\omega - i D_+) h_v(x) | \overline{B}(p_B)>
\eea

\noi with 
\bea
\label{55e}
&&<p_2,s_2| h_v(x)|\overline{B}(p_B)> = e^{i(p_2-p_B+m_bv)\cdot x} <p_2,s_2| h_v(0)|\overline{B}(p_B)> \nn \\
&&<B(p_B)| \overline{h}_v(x)|p_2,s_2> = e^{-i(p_2-p_B+m_bv)\cdot x} <\overline{B}(p_B)| \overline{h}_v(0)|p_2,s_2>
\eea

\noi We use now for the operator $\delta (\omega - iD_+)$ the following representation
\beq
\label{56e}
\delta (\omega - iD_+) = {1 \over 2 \pi} \int_{-\infty}^{\infty} ds\ e^{i(\omega - i D_+) s}
\eeq

\noi and make the identification in the quark model, that follows from the hypothesis (\ref{9e}) \beq
\label{57e}
D_+ \to \partial_+ = {\partial \over \partial x^+}
\eeq

\noi One obtains, for any function $f(x^+)$
\beq
\label{58e}
e^{s\partial_+} f(x^+) = \sum_{n=0}^{\infty} {1 \over n!} (s\partial_+)^n f(x^+) = f(x^+ + s)
\eeq

\noi from $p\cdot x =p_+x_- + p_- x_+ - {\bf p}_{\bot}\cdot {\bf x}_{\bot}$ with $p_{\pm} = p^0 \pm p^z$ and $\partial_+ = {\partial \over \partial x^+} = {\partial \over \partial x_-}$ one finds 
\bea
\label{59e}
&&e^{s\partial_+}\ e^{i(p_2 - p_B + m_b v )\cdot x}\nn\\
&&= e^{i \left \{ (p_2 - p_B + m_b v)_+ (x_- + s) + (p_2-p_B+m_b v)_- x_+ - ({\bf p}_2 - {\bf p}_B + m_b {\bf v})_{\bot} \cdot {\bf x}_{\bot} \right \} }
\eea

\noi and therefore
\beq
\label{60e}
\delta (\omega - i \partial_+) e^{i(p_2-p_B+m_bv)\cdot x} = \delta \left [\omega + (p_2 - p_B + m_b v)_+ \right ] e^{i(p_2-p_B+m_bv)\cdot x}
\eeq

\noi hence, in the quark model
\bea
\label{61e}
&&S(\omega ) \ {1 \over 2} \ Tr \left ( \Gamma {1 + {/ \hskip -2 truemm v} \over 2}\right ) = {<\overline{B}(p_B) | \overline{h}_v(x)  \delta (\omega - i \partial_+) h_v(x) | \overline{B}(p_B)> \over 2m_B}\nn \\
&&= {<\overline{B}(p_B) | \overline{h}_v(x)  \delta \left [ \omega +(p_2-p_B + m_b v)_+ \right ]  h_v(x) | \overline{B}(p_B)> \over 2m_B}
\eea

\noi In the $\overline{B}$ rest fame $m_bv_+ = m_b$, $p_B = (m_B, {\bf 0})$ and therefore, from $\overline{\Lambda} = m_B - m_b$ one gets
\beq
\label{62e}
S(\omega ) \ {1 \over 2} \ Tr \left ( \Gamma {1 + {/ \hskip -2 truemm v} \over 2}\right ) = {<\overline{B}(p_B) | \overline{h}_v(x)\Gamma  \delta (\omega + p_{2+} - \overline{\Lambda}) h_v(x) | \overline{B}(p_B)> \over 2m_B}
\eeq

One obtains for the shape function (\ref{52e})
\beq
\label{63e}
\widehat{S}(\widehat{\omega} ) \ {1 \over 2} \ Tr \left ( \Gamma {1 + {/ \hskip -2 truemm v} \over 2}\right ) = {<\overline{B}(p_B) | \overline{h}_v(x)\Gamma  \delta (\widehat{\omega} - p_{2+} )h_v(x) | B(p_B)> \over 2m_B}
\eeq

\noi In these expressions $p_{2+}$ is the component $p_{2+} = p_2^0 + p_2^z$ of the spectator light quark. \par

Let us now compute explicitly this last expression in the BT model, proceeding along the same lines as we have done in Section 6 for the LCDA. One needs to compute the matrix element of the operator $\Gamma \delta (\widehat{\omega} - p_{2+})$ in the forward direction
\bea
\label{64e}
&&<\overline{B}({\bf P})|\Gamma \delta (\widehat{\omega} - p_{2+})| \overline{B}({\bf P})> = \int {d{\bf p}_2 \over (2 \pi )^3} \ {1 \over p_2^0} \left [ \varphi ({\bf k}_2)\right ]^2 \delta (\omega - p_{2+})\\
&&{u^0 \over (p_1^0)}\ {k_1^0 k_2^0 \over \left ( k_1^0 + m_1\right ) \left ( k_2^0 + m_2\right )}\ {1 \over 16} Tr \left [ \Gamma (m_1 + {/ \hskip -2.5 truemm p}_1) (1 + {/ \hskip -2.5 truemm u}) (m_2 + {/ \hskip -2.5 truemm p}_2) (1 + {/ \hskip -2.5 truemm u}) (m_1 + {/ \hskip -2.5 truemm p}_1) \right ] \nn
\eea 

\noi where $k_i$ are related to $p_i$ through the boost (\ref{16e}), ${\bf B}_u k_i = p_i$. Changing the measure, and taking into account that now $p_{2+} = p_2^0 + p_2^z$, one can write
\bea
\label{66e}
&&<\overline{B}({\bf P})|\Gamma \delta (\widehat{\omega} - p_{2+}| \overline{B}({\bf P})> =\nn\\
&&\int  {dp_{2+} \over p_{2+}}\ {dp_{2-} \over 2 \pi} \ \delta \left ( p_{2-} - {{\bf p}_{\bot}^2 + m_2^2 \over p_{2+}}\right ) {d{\bf p}_{\bot} \over (2 \pi )^2}
{u^0 \over (p_1^0)}\ {k_1^0 k_2^0 \over \left ( k_1^0 + m_1\right )  \left ( k_2^0 + m_2\right )}\\
&& \left [ \varphi ({\bf k}_2)\right ]^2 \delta (\widehat{\omega} - p_{2+}){1 \over 16} Tr \left [ \Gamma (m_1 + {/ \hskip -2.5 truemm p}_1) (1 + {/ \hskip -2.5 truemm u}) (m_2 + {/ \hskip -2.5 truemm p}_2) (1 + {/ \hskip -2.5 truemm u}) (m_1 + {/ \hskip -2.5 truemm p}_1) \right ] \nn
\eea 

\noi Integrating relatively to $p_{2+}$ and $p_{2-}$
\bea
\label{67e}
&&<\overline{B}({\bf P}|\Gamma \delta (\widehat{\omega} - p_{2+})| \overline{B}({\bf P})> = {1 \over 16}\ {1 \over 2 \pi} \int {d{\bf p}_{\bot} \over (2 \pi )^2}\ {1 \over \widehat{\omega}}\nn\\
&& \left \{ \left [ \varphi ({\bf k}_2)\right ]^2{u^0 \over (p_1^0)^2}\ {k_1^0 k_2^0 \over \left ( k_1^0 + m_1\right ) \left ( k_2^0 + m_2\right )}\right . \\
&&\left . Tr \left [ \Gamma (m_1 + {/ \hskip -2.5 truemm p}_1) (1 + {/ \hskip -2.5 truemm u}) (m_2 + {/ \hskip -2.5 truemm p}_2) (1 + {/ \hskip -2.5 truemm u}) (m_1 + {/ \hskip -2.5 truemm p}_1) \right ] \right \}_{p_{2+} = \widehat{\omega}, p_{2-} = {{\bf p}_{\bot}^2 + m_2^2 \over  \hat{\omega}}}\nn
\eea 

\subsection{Heavy quark limit.}
\hspace*{\parindent}
In the heavy mass limit one has $u \to v_1 \to v$, $k_2^0 \to p_2\cdot v$ and $M_0 \to m_1 \to m_B$ and therefore, after some algebra,
\bea
\label{68e}
&&<\overline{B}({\bf P}|\Gamma \delta (\widehat{\omega} - p_{2+})| \overline{B}({\bf P})> = {1 \over 8}\ {1 \over 2 \pi} \int {d{\bf p}_{\bot} \over (2 \pi )^2}\ {1 \over \widehat{\omega}}\nn\\
&& \left \{ \left [ \varphi \left ( \sqrt{(p_2 \cdot v)^2 - m_2^2}\right ) \right ]^2 {1 \over v^0}\ {p_2 \cdot v \over p_2 \cdot v + m_2}\right . \\
&&\left . Tr \left [ \Gamma (1 + {/ \hskip -2.5 truemm v}) (m_2  + {/ \hskip -2.5 truemm p}_2 )  (1 + {/ \hskip -2.5 truemm v}) \right ] \right \}_{p_{2+} = \widehat{\omega}, p_{2-} = {{\bf p}_{\bot}^2 + m_2^2 \over  \hat{\omega}}}\nn
\eea 

\noi and from
\beq
\label{69e}
Tr \left [ \Gamma (1 + {/ \hskip -2.5 truemm v}) (m_2  + {/ \hskip -2.5 truemm p}_2 )  (1 + {/ \hskip -2.5 truemm v}) \right ]  = 8 \left [ m_2 + (p_2 \cdot v) \right ] {1 \over 2} Tr \left ( \Gamma {1 + {/ \hskip -2.5 truemm v}\over 2} \right )
\eeq

\noi one gets finally
\bea
\label{70e}
&&<\overline{B}({\bf P}|\Gamma \delta (\widehat{\omega} - p_{2+})| \overline{B}({\bf P}> = {1 \over 2}\ Tr \left ( \Gamma {1 + {/ \hskip -2.5 truemm v}\over 2} \right ) {1 \over 2 \pi} \int {d{\bf p}_{\bot} \over (2 \pi )^2}\ {1 \over \widehat{\omega}}\nn\\
&& \left \{ \left [ \varphi \left ( \sqrt{(p_2 \cdot v)^2 - m_2^2}\right ) \right ]^2 {1 \over v^0} (p_2 \cdot v)\right \}_{p_{2+} = \widehat{\omega}, p_{2-} = {{\bf p}_{\bot}^2 + m_2^2 \over  \hat{\omega}}}
\eea  

\noi Let us identify with the definition (\ref{52e}). With the normalisation of the BT model (\ref{20e}) we have to identify the matrix element (\ref{70e}) with
\beq
\label{71e}
{<\overline{B}({\bf P})|\Gamma \delta (\widehat{\omega} - p_{2+})| \overline{B}({\bf P})>  \over 2 M_B} = {1 \over 2} Tr \left ( \Gamma {1 + {/ \hskip -2.5 truemm v}\over 2} \right ) {\widehat{S}(\widehat{\omega}) \over 2P^0}
\eeq

\noi Therefore one gets
\beq
\label{72e}
\widehat{S}(\widehat{\omega}) =  {1 \over 2}\int {d{\bf p}_{\bot} \over (2 \pi )^2}\ {1 \over \omega }  \left \{ (p_2 \cdot v) \left [ \varphi \left ( \sqrt{(p_2 \cdot v)^2 - m_2^2}\right ) \right ]^2 \right \}_{p_{2+} = \widehat{\omega}, p_{2-} = {{\bf p}_{\bot}^2 + m_2^2 \over  \hat{\omega}}}
\eeq

\noi and from 
\beq
\label{73e}
p_2 \cdot v = {{\bf p}_{\bot}^2 + m_2^2 +  \widehat{\omega}^2 \over  2\widehat{\omega}}
\eeq

\noi one obtains finally
\beq
\label{74e}
\widehat{S}(\widehat{\omega}) =  {1 \over 2}\int {d{\bf p}_{\bot} \over (2 \pi )^2}\ {{\bf p}_{\bot}^2 + m_2^2 +  \widehat{\omega}^2 \over  2\widehat{\omega}} \left [ \varphi \left ( {\sqrt{({\bf p}_{\bot}^2  +  \widehat{\omega}^2 + m_2^2)^2 - 4  \widehat{\omega}^2 m_2^2}   \over  2\widehat{\omega}}\right ) \right ]^2
\eeq

\noi Again, similarly to the LCDA, one can obtain a more compact form of $\widehat{S}(\widehat{\omega})$ by peforming the change of variables
\beq
\label{75e}
k = {\sqrt{({\bf p}_{\bot}^2  +  \widehat{\omega}^2 + m_2^2)^2 - 4  \widehat{\omega}^2 m_2^2}   \over  2\widehat{\omega}}
\eeq

\noi and one obtains the simple final result for the Shape Function in the BT model
\beq
\label{76e}
\widehat{S}(\widehat{\omega}) =  {1 \over 4\pi^2} \int_{k_0 (\widehat{\omega})}^{\infty} dk\ k [\varphi (k)]^2 \qquad (0 \leq \widehat{\omega} < \infty ) 
\eeq

\noi with
\beq
\label{77e}
k_0 (\widehat{\omega}) = {|\widehat{\omega} - m_2|(\widehat{\omega} + m_2) \over 2\widehat{\omega}}
\eeq

\noi that gives the value $\widehat{\omega}_{max}$ for which $\widehat{S}(\widehat{\omega})$ attains its maximum value $\widehat{S}_{max} = \widehat{S}(\widehat{\omega}_{max})$,
\beq
\label{78e}
\widehat{\omega}_{max} = m_2 \qquad \qquad \widehat{S}_{max} = {1 \over 4 \pi^2} \int_0^{\infty}  dk\ k [\varphi (k)]^2 
\eeq

For the other definition of the shape function $S( \omega )$ (\ref{50e}) one reads, from (\ref{52e}),
\beq
\label{79e}
S( \omega ) = {1 \over 4 \pi^2} \int_{\kappa_0(\omega )}^{\infty} dk\ k [\varphi (k)]^2 
\eeq

\noi with
\beq
\label{80e}
\kappa_0 (\omega ) = {|\overline{\Lambda} - \omega - m_2| (\overline{\Lambda} - \omega + m_2) \over 2 (\overline{\Lambda} - \omega)} \qquad (- \infty < \omega \leq \overline{\Lambda})
\eeq

To check the normalization condition
\beq
\label{81e}
\int_0^{\infty} d \widehat{\omega}\ \widehat{S}(\widehat{\omega}) = 1
\eeq

\noi we use (\ref{79e}) and exchange the order of the integrals over $\widehat{\omega}$ and $k$,
\beq
\label{82e}
\widehat{\omega} > 0\ , \quad k > k_0 (\widehat{\omega})\qquad  \Leftrightarrow \qquad k > 0\ , \quad  \widehat{\omega}_-(k) < \widehat{\omega} < \widehat{\omega}_+(k)
\eeq

\noi where
\beq
\label{83e}
\widehat{\omega}_{\pm}(k) = \sqrt{k^2 + m_2^2} \pm k
\eeq

\noi One obtains
\beq
\label{84e}
\int_0^{\infty} d\widehat{\omega} \ \widehat{S}(\widehat{\omega}) = {1 \over 4 \pi^2} \int_0^{\infty} dk\ k [\varphi (k)]^2 \int_{\widehat{\omega}_-(k)}^{\widehat{\omega}_+(k)} d\widehat{\omega} = {1 \over 2 \pi^2} \int_0^{\infty} dk\  k^2 [\varphi (k)]^2 = 1
\eeq

\noi from the normalization of the internal wave function (\ref{15e}).\par

From (\ref{84e}) and (\ref{52e}) it follows, for the function $S(\omega )$,
\beq
\label{85e}
\int_{-\infty}^{\overline{\Lambda}} d \omega \ S( \omega ) = 1
\eeq

\subsection{Gaussian wave function.}
\hspace*{\parindent}
In principle, one could calculate the wave function $\varphi (k)$ entering in $\varphi_{\pm}^B(\omega )$ (\ref{34e}) and $S(\omega )$ (\ref{79e}) from the quark potential. However, as argued in the Introduction, we prefer here to adopt simply a gaussian wave function, that has been used elsewhere to obtain $S(\omega )$ and to fit the $B \to X_s \gamma$ spectrum \cite{20r,21r,22r,26r}.\par

Let us compute the shape function $S (\omega )$ using the harmonic oscillator potential, with the internal wave function at rest $\varphi ({\bf k}_2)$ is given by
\beq
\label{86e}
\varphi_{HO}({\bf k}_2) = (2 \pi )^{3/2} \left ( {R^2 \over \pi} \right )^{3/4} \exp \left ( - {R^2 {\bf k}_2^2 \over 2}\right )  
\eeq

\noi We obtain
\beq
\label{87e}
S( \omega ) = {R \over \sqrt{\pi}} \exp \left \{ - {1 \over 4}\ {\left [ ( \overline{\Lambda} -  \omega )^2 - m_2^2 \right ]^2 \over R^2 (\overline{\Lambda} - \omega )^2} \right \}
\eeq

\noi With the notation
\beq
\label{88e}
R = {1 \over p_F}\qquad\qquad \rho = {m_2^2 \over p_F^2} \qquad \qquad x = {\omega \over \overline{\Lambda}}
\eeq

\noi one gets 
\beq
\label{89e}
S( \omega ) = {1 \over \sqrt{\pi}} \ {1 \over p_F} \exp \left \{ - {1 \over 4}\ \left [ {p_F \over \overline{\Lambda} }\ {\rho \over 1 - x} - {\overline{\Lambda} \over p_F} (1 - x) \right ]^2 \right \} \qquad \left ( x = {\omega \over \overline{\Lambda} }\right )
\eeq

\noi that is correctly normalized to 1. \par

We obtain therefore in the BT model with harmonic oscillator potential the so-called ``Roman'' Shape Function \cite{20r,21r,22r,26r}.

\section{Predictions for the LCDA $\varphi_\pm^B(\omega )$.}
\subsection{Parameters of the Roman Shape Function.}
\hspace*{\parindent}
Limosani and Nozaki \cite{26r} have made a recent fit to the Belle $B \to X_s \gamma$ data \cite{32r}, in order to extract the $b$-quark shape function parameters $\overline{\Lambda}_{SF}$ and $\lambda_1^{SF}$ ($\lambda_1^{SF} = - \mu_{\pi}^2)$ in a number of models for the shape function, among them the Roman Shape Function. Writing (\ref{88e}) in their notation
\beq
\label{90e}
\omega = k_+ \qquad\qquad \qquad F(k_+) = S(\omega )
\eeq
\beq
\label{91e}
F(k_+) = N {\kappa \over \sqrt{\pi}} \exp \left \{ - {1 \over 4} \left [ {1 \over \kappa } \ {\rho \over 1 - x} - \kappa (1 - x) \right ]^2 \right \}
\eeq

\noi with
\beq
\label{92e}
x = {k_+ \over \overline{\Lambda}_{SF}} \qquad\qquad  \kappa = {\overline{\Lambda}_{SF} \over p_F} \qquad\qquad   \rho = {m_2^2 \over p_F^2} 
\eeq

\noi amounts to replace in (\ref{89e}) $\overline{\Lambda}$ by $\overline{\Lambda}_{SF}$ of the Shape Function renormalization scheme \cite{33r}. Of course, the BT quark model is not field theory, and this scheme is too rough to distinguish between $\overline{\Lambda}$ and $\overline{\Lambda}_{SF}$, that differ by QCD radiative corrections. Therefore, we can make the replacement of $\overline{\Lambda}$ by $\overline{\Lambda}_{SF}$ in the formulas of the precedent section, and this will be our final model for the shape function. \par

The first moments of $F(k_+)$
\beq
\label{93e}
A_n = \int_{-\infty}^{\overline{\Lambda}_{SF}} dk_+\ k_+^n \ F(k_+)
\eeq

\noi are given in QCD by
\beq
\label{94e}
A_0 = 1 \qquad\qquad  A_1 = 0\qquad\qquad A_2 = - {\lambda_1^{SF} \over 3}
\eeq

\noi Computing them explicitly using the shape function (\ref{91e}) one obtains
\beq
\label{95e}
A_n = N (\overline{\Lambda}_{SF})^{n+1} {\kappa \over \sqrt{\pi}}\ e^{\rho /2} \sum_k (-1)^k  {n \choose k} \left ( {\rho \over \kappa^2}\right ) ^{{k+1 \over 2}}\ K_{{k+1 \over 2}} \left ( {\rho \over 2}\right )
\eeq

\noi and the conditions (\ref{94e}) yield respectively
\bea
\label{96e}
&&N = {1 \over \overline{\Lambda}_{SF}} \nn \\
&&\kappa = {\rho \over \sqrt{\pi}}\ e^{\rho /2}\ K_1 \left ( {\rho \over 2}\right )\nn \\
&&\lambda_1^{SF} = - 3 \left [ {2 \over \kappa^2} \left ( {1 + \rho \over 2}\right ) - 1 \right ] (\overline{\Lambda}_{SF})^2
\eea

The best fit of Limosani and Nozaki \cite{26r} gives
\beq
\label{97e}
\overline{\Lambda}_{SF} = 0.66\ {\rm GeV} \qquad\qquad \lambda_1^{SF} = - 0.39\ {\rm GeV}^2
\eeq

\noi that corresponds, using (\ref{96e}), to the values
\beq
\label{98e}
\rho = 0.776 \qquad\qquad\qquad  \kappa = 1.462
\eeq

\noi or, from (\ref{88e}) and (\ref{92e}), to the quark model parameters
\beq
\label{99e}
R = {1 \over p_F} = 2.216 \ {\rm GeV}^{-1} \qquad \qquad m_2 = 0.398\ {\rm GeV}
\eeq

\noi On can also compute the value $(k_+)_{max}$ of $k_+$, for which $F(k_+)$ becomes maximum, $F_{max} = F[(k_+)_{max}]$,
\beq
\label{100e}
(k_+)_{max} = \left ( 1 - {\sqrt{\rho} \over \kappa}\right )\overline{\Lambda}_{SF} \qquad \qquad F_{max} = N \ {\kappa \over \sqrt{\pi}}
\eeq

\noi or numerically,
\beq
\label{101e}
(k_+)_{max} = 0.262  \qquad \qquad F_{max} = 1.250\ {\rm GeV}^{-1}
\eeq

The other definition of the Roman shape function $\widehat{S} (\widehat{\omega })$ (\ref{52e}) has a very simple expression in the BT model
\beq
\label{102e}
\widehat{S} (\widehat{\omega }) = {R \over \sqrt{\pi}} \ \exp \left [ - {R^2 \over 4} \left ( {m_2^2 \over \widehat{\omega }} -\widehat{\omega }\right )^2 \right ]
\eeq

\noi that is identical to $\widehat{S} (\widehat{\omega })_{Roman}$ with 
\bea
\label{103e}
&&\rho = R^2 m_2^2 \nn \\
&&\overline{\Lambda}_{SF} = {R^2 m_2^2 \over \sqrt{\pi}} \ \exp \left ( {R^2m_2^2 \over 2}\right ) K_1 \left ( {R^2m_2^2 \over 2}\right )\nn \\
&&\lambda_1^{SF} = - 3 \left [ {2 \over R^2} + m_2^2 - (\overline{\Lambda}_{SF} )^2 \right ]
\eea

\noi and
\beq
\label{104e}
\widehat{\omega}_{max} = 0.398\ {\rm GeV} \qquad \qquad \widehat{S}_{max} = 1.250\ {\rm GeV}^{-1} 
\eeq
 
 \subsection{Predictions for $\varphi_{\pm}^B(\omega )$.}
\hspace*{\parindent}
Using the harmonic oscillator wave function (\ref{86e}), the parameters (\ref{99e}) and expressions (\ref{31e}) or (\ref{34e}) we can predict the LCDA $\varphi_\pm^B(\omega )$ within the BT model.\par

In the limit of vanishing light quark mass $m_2 = 0$, in which the relation (\ref{39e}) between $\varphi_+^B(\omega )$ and $\varphi_-^B(\omega )$ holds, one obtains simple analytic expressions in terms of the Error Function $\Phi (x)$
\beq
\label{105e}
\Phi (x) = {2 \over \sqrt{\pi}} \int_0^x dt\ e^{-t^2}
\eeq

\noi For the decay constant one gets
\beq
\label{106e}
f_B \sqrt{m_B} = \sqrt{N_c}\ {1 \over \pi^{3/4}} \sqrt{{R \over 2}} \int_0^{\infty} d \omega\ \omega \left [ 1 - \Phi \left ({R\omega \over 2 \sqrt{2}}\right ) \right ]
\eeq

\noi and for the LCDA, 
\bea
\label{107e}
&&\varphi_+^B(\omega ) = { \sqrt{N_c} \over f_B \sqrt{m_B}} \ {1 \over \pi^{3/4}} \sqrt{{R \over 2}}  \omega \left [ 1 - \Phi \left ({R\omega \over 2 \sqrt{2}}\right ) \right ] \\
&&\varphi_-^B(\omega ) = { \sqrt{N_c} \over f_B \sqrt{m_B}} \ {1 \over \pi^{3/4}} \sqrt{{2 \over R}} \left \{ \sqrt{{2 \over \pi}} \exp \left ( - {R^2 \omega^2 \over 8}\right ) - {R \omega \over 2}  \left [ 1 - \Phi \left ({R\omega \over 2 \sqrt{2}}\right ) \right ]\right \}\nn \eea

To illustrate numerical results, and give a feeling of the dependence on the light quark mass, we chose two sets of parameters~:\par

(1) The realistic case of the parameters of the quark wave function (\ref{99e}) $R = 2.216$~GeV$^{-1}$ and $m_2 = 0.398$~GeV. From these parameters fitted on the $B \to X_s \gamma$ spectrum we predict 
\beq
\label{108NEWEq}
f_B \sqrt{m_B} = 0.388 \hbox{ GeV$^{3/2}$}
\eeq 

\noi and the functions $\varphi_+^B(\omega )$ and $\varphi_-^B(\omega )$ that are plotted respectively in Figs.~1 and 2. The heavy quark limit value (\ref{108NEWEq}) gives, using the physical $B$ mass, $f_B \cong 170$~MeV, a little smaller than the popular values for this quantity. \par

(2) For comparison we adopt the vanishing light quark mass case, taking the same radius $R = 2.216$~GeV$^{-1}$ but $m_2 = 0$. We find $f_B \sqrt{m_B} = 0.315$~GeV$^{3/2}$ and the functions $\varphi_+^B(\omega )$ and $\varphi_-^B(\omega )$ that are also plotted in Figs.~1 and 2.\par

\begin{figure}
\centering\leavevmode
\includegraphics[scale=1.4]{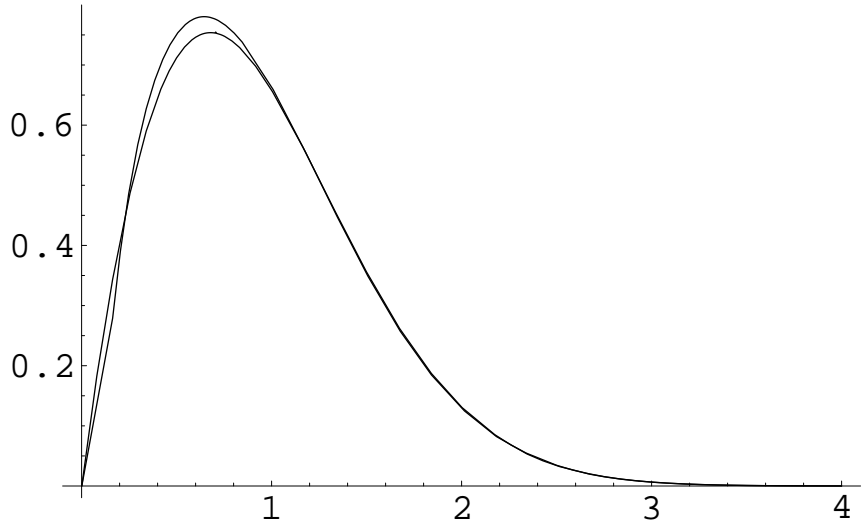}
\caption{The function $\varphi_+^B(\omega )$ for a gaussian wave function for the sets of parameters \protect{(\ref{99e})} obtained from the Roman Shape Function \protect{\cite{26r}} $(R,m_2) = (2.216$~GeV$^{-1}$, 0.398~GeV) (higher curve) and $(R, m_2) = (2.216$~GeV$^{-1}$, 0) (lower curve). For non-vanishing $m_2$ the first derivative vanishes at $\omega = 0$.
\centerline{\hbox to 8 truecm{\hrulefill}}}
\label{fig.1}
\end{figure}
 
\begin{figure}
\centering\leavevmode
\includegraphics[scale=1.7]{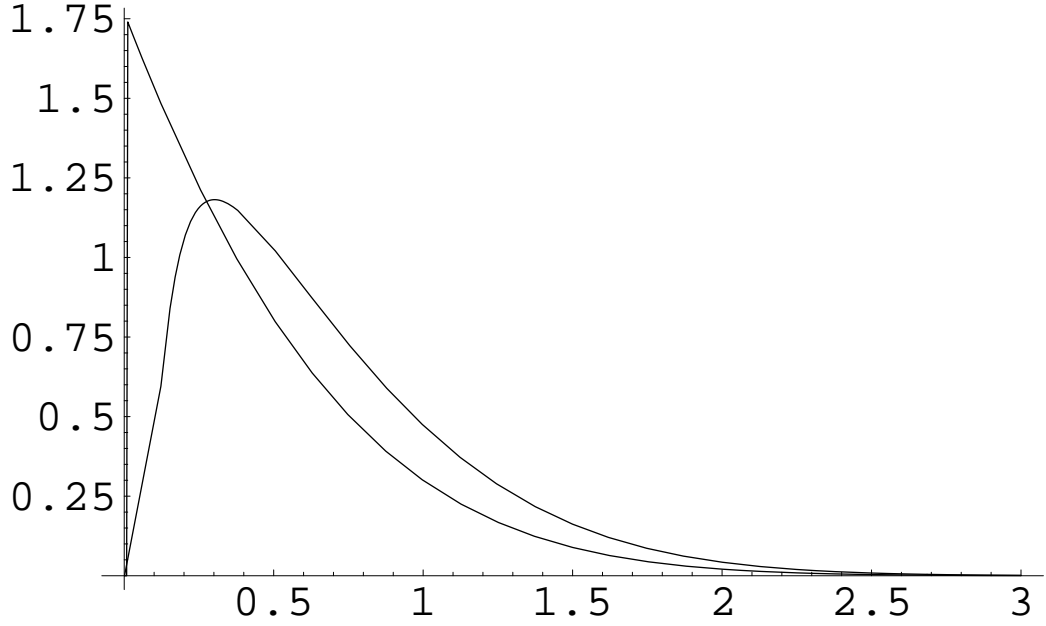}
\caption{The function $\varphi_-^B(\omega )$ for a gaussian wave function for the sets of parameters obtained from the Roman Shape Function \protect{\cite{26r}} $(R,m_2) = (2.216$~GeV$^{-1}$, 0.398~GeV) and $(R, m_2) = (2.216$~GeV$^{-1}$, 0). For non-vanishing $m_2$ the function vanishes at $\omega = 0$.
%\centerline{\hbox to 8 truecm{\hrulefill}}
}
\label{fig.2}
\end{figure}

We observe that the non-vanishing light quark mass gives $\varphi_-^B(0) = 0$ and $\varphi_+^{B'} (0) = 0$, while for vanishing light quark mass one obtains $\varphi_-^B(0) \not= 0$ and also $\varphi_+^{B'} (0) \not= 0$ like in QCD Sum rules. Since the vanishing $\varphi_-^B(0) = 0$ for $m_2 \not= 0$ is an unfamiliar feature, we plot in Fig.~3 the evolution of $\varphi_-^B (\omega )$ with increasing values of $m_2 = 0$, $0.1$, $0.2$, $0.3$~GeV.

\begin{figure}
\centering\leavevmode
\includegraphics[scale=1.7]{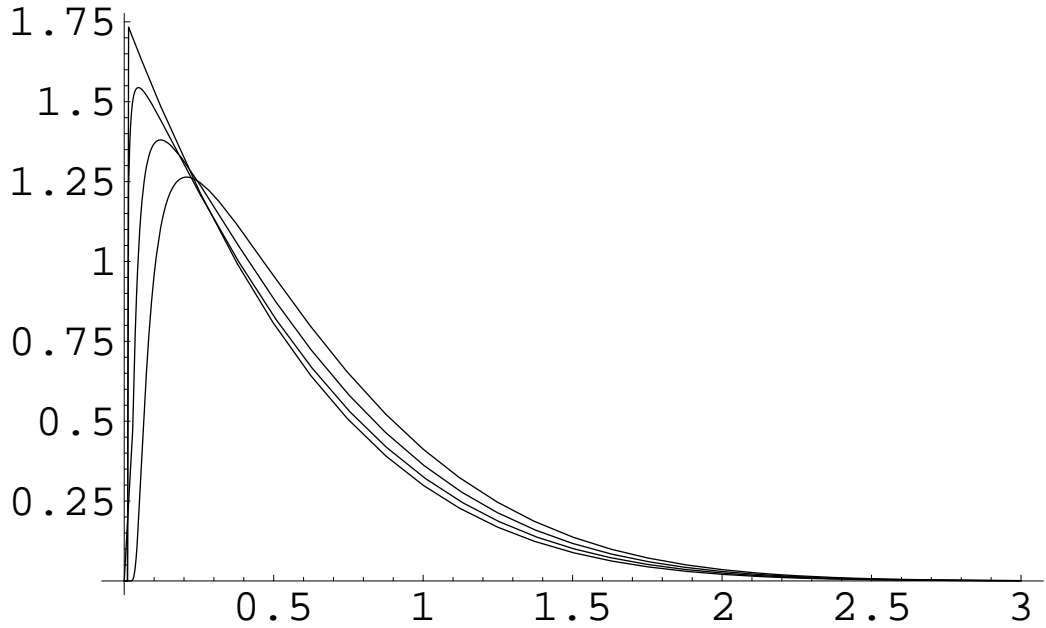}
\caption{Evolution of $\varphi_-^B(\omega )$ with the constituent light quark mass for $R = 2.216$~GeV$^{-1}$ and $m_2 = 0$, $0.1$, $0.2$, $0.3$~GeV.
%\centerline{\hbox to 8 truecm{\hrulefill}}
}
\label{fig.3}
\end{figure}

\subsection{Moments of $\varphi_{+}^B(\omega )$.}
\hspace*{\parindent}
Let us now compute the moments of $\varphi_{\pm}^B(\omega )$ using the harmonic oscillator wave function (\ref{86e}) that gives the Roman Shape Function (\ref{91e}),  (\ref{92e}) with the parameters (\ref{99e}). We obtain for the first moments $M_{\pm}^{(1)}$
\beq
\label{108e}
M_+^{(1)BT} = 2 M_-^{(1)BT} \cong {4 \overline{\Lambda}_{SF}\over 3} = 0.964\ {\rm GeV}
\eeq

\noi that gives $\overline{\Lambda}_{SF} \cong 0.723$~GeV, in qualitative agreement at the 10\% level with the value (\ref{97e}) obtained from the fit to $B \to X_s \gamma$, and for the moment $M_+^{(-1)BT}$, called also $\lambda_B^{-1}$
\beq
\label{109e}
\left ( \lambda_B^{-1}\right )^{BT} = M_+^{(-1)BT} = 1.521\ {\rm GeV}^{-1} 
\eeq

\section{Radiative corrections.}
\hspace*{\parindent}
The importance of the radiative tail has been underlined by Braun, Ivanov and Korchemsky \cite{7r}. When taking into account one-loop QCD corrections, the moments for $N \geq 0$ of $\varphi_{+}^B(\omega )$ are divergent, as already pointed out by Grozin and Neubert \cite{2r} and by Descotes-Genon and Sachrajda \cite{6r}.\par

Braun et al. give a parametrization of $\varphi_{+}^B(\omega , \mu )$ based on QCD Sum Rules plus the QCD behaviour $- {1 \over \omega} {\rm Log} \left ( {\omega \over \mu}\right )$ at large $\omega$, 
\beq
\label{110e}
\varphi_{+}^B(\omega , \mu ) = {4 \lambda_B^{-1} \over \pi} \ {\omega \mu \over \omega^2 + \mu^2} \left [ {\mu^2 \over \omega^2 + \mu^2} - {2(\sigma_B - 1) \over \pi^2} \ {\rm Log} \left ( {\omega \over \mu}\right ) \right ]
\eeq

\noi where the parameters $\lambda_B^{-1}$ and $\sigma_B$ are defined in terms of the integrals
\bea
\label{111e}
&&\lambda_B^{-1} = \int_0^{\infty} d\omega {\varphi_+^B (\omega , \mu ) \over \omega } \nn \\
&&\sigma_B \lambda_B^{-1} = - \int_0^{\infty} d\omega {\varphi_+^B (\omega , \mu ) \over \omega }  \ {\rm Log} \left ( {\omega \over \mu}\right ) 
\eea

\noi We realize that, due to the logarithm in the radiative tail, the moments $M_+^{(N)}$ for $N \geq 0$ are divergent. \par

As explained below in Subsection 10.11, the parametrization (\ref{110e}) is a simplified and approximated form of a full QCD sum rules calculation that includes radiative corrections.\par

A thorough study of the radiative corrections has been done by Lee and Neubert \cite{9r}. The lowest positive moments $M_+^{(0)}$ and $M_+^{(1)}$ are given as a power expansion in the ultraviolet cut-off $\Lambda_{UV}$
\bea
\label{112e}
&&M_+^{(0)}(\Lambda_{UV}) = 1 + {C_F \alpha_s \over 4 \pi} \left [ - 2\  {\rm Log}^2 \left ( {\Lambda_{UV} \over \mu}\right ) + 2\  {\rm Log} \left ({\Lambda_{UV} \over \mu}\right ) - {\pi^2 \over 12}\right ] \nn \\
&&+ {16  \overline{\Lambda} \over 3\Lambda_{UV}}\ {C_F \alpha_s \over 4 \pi} \left [ {\rm Log} \left ({\Lambda_{UV} \over \mu}\right ) - 1 \right ]
\eea
\bea
\label{113e}
&&M_+^{(1)}(\Lambda_{UV}) =  \Lambda_{UV} {C_F \alpha_s \over 4 \pi} \left [ - 4\  {\rm Log} \left ({\Lambda_{UV} \over \mu}\right ) + 6\  \right  ] \nn \\
&&+ {4  \overline{\Lambda} \over 3} \left \{ 1 + {C_F \alpha_s \over 4 \pi} \left [ - 2 {\rm Log}^2 \left ({\Lambda_{UV} \over \mu}\right ) + 8  \  {\rm Log} \left ( {\Lambda_{UV} \over \mu} \right ) - {7 \over 4} - {\pi^2 \over 12}\right ] \right \} 
\eea

\noi In the limit $\alpha_s \to 0$, one recovers $M_0 = 1$ and $M_1 = {4   \overline{\Lambda}\over 3}$. Lee and Neubert define the radiative tail of the function $\varphi_+^B (\omega , \mu )$ by the prescription 
\beq
\label{114e}
\varphi_+^{RAD} (\omega  ) = \left [ {dM_0(\Lambda_{UV}, \mu) \over d \Lambda_{UV}}\right ]_{\Lambda_{UV} = \omega}
\eeq

\noi that gives
\beq
\label{115e}
\varphi_+^{RAD} (\omega  ) = {C_F \alpha_s \over  \pi\omega } \left \{ \left [ {1 \over 2} - \ {\rm Log } \left ( {\omega \over \mu}\right )\right ]  + {4  \overline{\Lambda} \over 3}\ {1 \over \omega} \left [ 2 - \ {\rm Log } \left ( {\omega \over \mu}\right ) \right ] \right \} 
\eeq

\noi Notice that for large enough $\omega$ the radiative tail becomes negative, and that at lowest order it agrees with (\ref{110e}). \par

To include the radiative tail is not without ambiguity. To add it to our calculation of the long distance part of $\varphi_{+}^B(\omega )$, we follow two different models, that have different continuity properties, but that lead to almost identical results~: first, a similar procedure to the one proposed by Lee et al.  \cite{9r}, although different in its details; second, a procedure close to the one followed by Braun et al. \cite{7r}. We now expose both methods, and compare the results at the end.

\subsection{Model to add the radiative tail following Lee et al.}
\hspace*{\parindent}
Our first model for the function $\varphi_{+}^B(\omega )$, including the radiative corrections follows essentially the prescription of \cite{9r}
\beq
\label{116e}
\varphi_{+}^B(\omega ) = N \varphi_{+}^{BT}(\omega ) + \theta (\omega - \omega_t ) \varphi_+^{RAD}  (\omega  ) 
\eeq

\noi where $ \varphi_{+}^{BT}(\omega )$ is the function (\ref{31e}) or (\ref{34e}) with the harmonic oscillator wave function (\ref{86e}) and the parameters of the Roman shape function (\ref{99e}).\par

Without loss of generality at leading order in $\alpha_s$, we take the radiative tail $\varphi_{+}^{RAD}(\omega )$ as given by expression (\ref{115e}) with $\overline{\Lambda}$ replaced now by $\overline{\Lambda}_{SF}(\mu )$, as determined from the fit to $B \to X_s \gamma$ 
\beq
\label{117e}
\varphi_+^{RAD} (\omega  ) = {C_F \alpha_s \over  \pi\omega } \left \{ \left [ {1 \over 2} - \ {\rm Log } \left ( {\omega \over \mu}\right ) \right ] + {4  \overline{\Lambda}_{SF}(\mu ) \over 3}\ {1 \over \omega} \left [ 2 - \ {\rm Log } \left ( {\omega \over \mu}\right ) \right ] \right \} 
\eeq

\noi The relation between $\overline{\Lambda}_{SF}(\mu ) \equiv \overline{\Lambda}_{SF}(\mu , \mu )$ and $\overline{\Lambda}$ is \cite{25r}
\beq
\label{118e}
\overline{\Lambda} = \overline{\Lambda}_{SF}(\mu )  - {C_F \alpha_s \over 4 \pi} \ 4 \mu
\eeq

We choose from now on $\mu = 1.5$~GeV as an illustration, 
\beq
\label{119e}
\mu = 1.5\ {\rm GeV} \qquad\qquad C_F \alpha_s = 0.470
\eeq

\noi For this value of $\mu$ one has, from $B \to X_s \gamma$ and $B \to X_u\ell \overline{\nu}_{\ell}$ \cite{9r,33r},
\beq
\label{120e}
\overline{\Lambda}_{SF}(\mu )  = (0.65 \pm 0.06)\ {\rm GeV} \qquad (\mu = 1.5\ {\rm GeV})
\eeq

\noi that is in agreement with the determination (\ref{97e}) from \cite{26r}. 

The relation between $\omega_t$, defined by the vanishing of the radiative tail to ensure continuity, and $\overline{\Lambda}_{SF}(\mu )$ is given by 
\beq
\label{122e}
{1 \over 2} - \ {\rm log} \left ( {\omega_t \over \mu}\right ) + {4\overline{\Lambda}_{SF}(\mu )  \over 3} \ {1 \over \omega_t} \left [ 2 - \ {\rm log} \left ( {\omega_t \over \mu}\right ) \right ] = 0
\eeq

\noi $\overline{\Lambda}_{SF}(\mu )$ becomes a function of $\omega_t$ and $N$ is a parameter to be determined by the matching with the QCD behaviour (\ref{112e}), (\ref{113e}).\par

In terms of $\overline{\Lambda}_{SF}(\mu )$, at the lowest order in $\alpha_s$ one can rewrite the moments (\ref{112e}) and (\ref{113e})
\bea
\label{123e}
&&M_+^{(0)}(\Lambda_{UV}) = 1 + {C_F \alpha_s \over 4 \pi} \left [ - 2\  {\rm Log}^2 \left ( {\Lambda_{UV} \over \mu}\right ) + 2\  {\rm Log} \left ({\Lambda_{UV} \over \mu}\right ) - {\pi^2 \over 12}\right ] \nn \\
&&+ {16  \overline{\Lambda}_{SF}(\mu ) \over 3\Lambda_{UV}}\ {C_F \alpha_s \over 4 \pi} \left [ {\rm Log} \left ({\Lambda_{UV} \over \mu}\right ) - 1 \right ]
\eea
\bea
\label{124e}
&&M_+^{(1)}(\Lambda_{UV}) =  \Lambda_{UV} {C_F \alpha_s \over 4 \pi} \left [ - 4\  {\rm Log} \left ({\Lambda_{UV} \over \mu}\right ) + 6\  \right  ] \nn \\
&&+ {4  \overline{\Lambda}_{SF}(\mu ) \over 3} \left \{ 1 + {C_F \alpha_s \over 4 \pi} \left [ - 2 {\rm Log}^2 \left ({\Lambda_{UV} \over \mu}\right ) + 8  \  {\rm Log} \left ( {\Lambda_{UV} \over \mu}\right ) - {7 \over 4} - {\pi^2 \over 12}\right ] \right \}\nn \\
&&-  {C_F \alpha_s \over 4 \pi}\ {16 \mu \over 3}
\eea

\noi Notice that a new term proportional to $\alpha_s$ appears in the first moment $M_+^{(1)}(\Lambda_{UV})$. \par

Using the LCDA $\varphi_{+}^B(\omega )$ from the BT model (\ref{34e}) with the harmonic oscillator wave function (\ref{86e}) and the parameters (\ref{99e}) we then compute the moments in the model
\bea
\label{125e}
&&M_+^{(0)model}(\Lambda_{UV}) = \int_0^{\Lambda_{UV}} d\omega \ \varphi_{+}^B(\omega )\nn \\
&&M_+^{(1)model}(\Lambda_{UV}) = \int_0^{\Lambda_{UV}} d\omega \ \omega \ \varphi_{+}^B(\omega )
\eea

\noi and match with the OPE expressions (\ref{123e}) and (\ref{124e}).\par

Making the approximation, that will be discussed below,
\bea
\label{126e}
&& \int_0^{\Lambda_{UV}} d\omega \ \varphi_{+}^B(\omega ) \cong \int_0^{\infty} d\omega \ \varphi_{+}^B(\omega ) \\
&&\int_0^{\Lambda_{UV}} d\omega \ \omega \ \varphi_{+}^B(\omega )  \cong \int_0^{\infty} d\omega \ \omega \ \varphi_{+}^B(\omega )
\label{127e}
\eea

\noi the matching implies
\bea
\label{128e}
&&M_+^{(0)}(\Lambda_{UV}) = N\ M_+^{(0)BT} + M_+^{(0)}(\Lambda_{UV}) - M_+^{(0)}(\omega_t) \nn \\
&&M_+^{(1)}(\Lambda_{UV}) = N\ M_+^{(1)BT} + M_+^{(1)}(\Lambda_{UV}) - M_+^{(1)}(\omega_t)
\eea

\noi that gives, since $M_+^{(0)BT} = 1$,
\bea
\label{129e}
&&N =  M_+^{(0)}(\omega_t)   \\
&&M_+^{(1)BT} = {M_+^{(1)}(\omega_t) \over M_+^{(0)}(\omega_t)} = {M_+^{(1)}(\omega_t) \over N}
\label{130e}
\eea

Equation (\ref{129e}) gives $N$ in terms of $\omega_t$
\bea
\label{131e}
&&N(\omega_t, \mu ) = 1 + {C_F \alpha_s \over 4 \pi} \left [ - 2\  {\rm Log}^2 \left ( {\omega_t \over \mu}\right ) + 2\  {\rm Log} \left ({\omega_t \over \mu}\right ) - {\pi^2 \over 12}\right ] \nn \\
&&+ \ {16  \overline{\Lambda}_{SF}(\mu ) \over 3\omega_t } \ {C_F \alpha_s \over 4 \pi} \left [ {\rm Log} \left ({\omega_t \over \mu}\right ) - 1 \right ]
\eea

and equation (\ref{130e}) gives, expanding to first order in $\alpha_s$,
\bea
\label{132e}
&&M_+^{(1)BT} =  {4  \overline{\Lambda}_{SF}(\mu ) \over 3} \left [ 1 + {C_F \alpha_s \over 4 \pi} \left \{ -6\  {\rm Log} \left ( {\omega_t \over \mu}\right ) - {16  \overline{\Lambda}_{SF}(\mu ) \over 3\omega_t }  \left [ {\rm Log} \left ({\omega_t \over \mu}\right ) - 1 \right ] \right \} \right ]\nn \\
&& - {C_F \alpha_s \over 4 \pi} \ \omega_t \left [ 4\ {\rm Log} \left ({\omega_t \over \mu}\right ) - 6 \right ] -  {C_F \alpha_s \over 4 \pi}\ {16 \mu \over 3}
\eea

Using the result (\ref{108e}) of the model $M_+^{(1)BT} = 0.964$~GeV, this equation gives another relation, besides (\ref{122e}), relating $\omega_t$ and $\overline{\Lambda}_{SF}(\mu )$. \par

From (\ref{122e}) and (\ref{132e}) we can solve for $\omega_t$ and $\overline{\Lambda}_{SF}(\mu )$ for a given value of $\mu$ and see if the value obtained for $\overline{\Lambda}_{SF}(\mu )$ is consistent with the known value of $\overline{\Lambda}_{SF}(\mu )$ (\ref{120e}) from the fit to $B \to X_s \gamma$. Once $\omega_t$ and $\overline{\Lambda}_{SF}(\mu )$ are known one can compute $N$ and $\overline{\Lambda}$ from (\ref{131e}) and (\ref{118e}).\par

From (\ref{122e}) and (\ref{132e}) we obtain, for $\mu = 1.5$~GeV,
\bea
\label{133e}
&\omega_t = 3.288 \ {\rm GeV} \qquad &\qquad\qquad N = 0.974 \nn \\
&\overline{\Lambda}_{SF}(\mu ) = 0.578\ {\rm GeV} &\qquad\qquad \overline{\Lambda} = 0.354\ {\rm GeV}
\eea

\noi These values for $\overline{\Lambda}_{SF}(\mu )$ and $\overline{\Lambda}$ that come from the OPE constraints are only about 10 \% lower from the values (\ref{120e}) coming from the fit to $B \to X_s \gamma$ and $B \to X_u \ell \overline{\nu}_{\ell}$. We conclude that the situation is good enough.\par

A different status from $\overline{\Lambda}$ has the parameter $\lambda_B^{-1}$, that enters in a number of processes that we examine blow. The value that we obtain including the radiative corrections is
\beq
\label{134e}
\lambda_B^{-1} = 1.429\ {\rm GeV}^{-1}
\eeq

\noi to be compared with the value (\ref{109e}) without the radiative tail. The correction is small. We find, for the parameter $\sigma_B$ (\ref{111e})
\beq
\label{135e}
\sigma_B = 1.207
\eeq

\noi The function (\ref{116e}) with its radiative tail is plotted in Fig. 4. 
\begin{figure}
\centering\leavevmode
\includegraphics[scale=1.4]{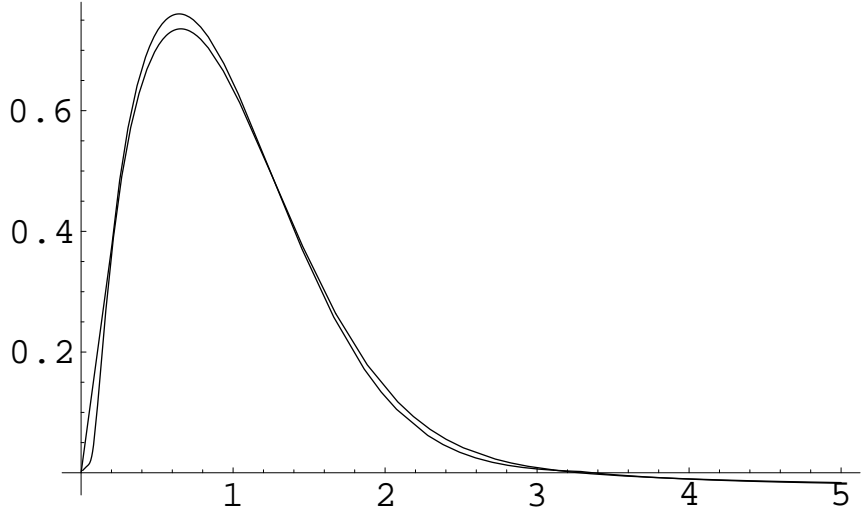}
\caption{The function $\varphi_+^B(\omega )$ in the BT approach with its radiative tail adopting the SF renormalization scheme for $\mu = 1.5$~GeV. The two curves show the two ways of gluing the radiative tail, following Lee et al. \protect{\cite{9r}} (lower curve) or Braun et al. \protect{\cite{7r}} (upper curve).
\centerline{\hbox to 8 truecm{\hrulefill}}}
\label{fig.4}
\end{figure}

\subsection{Model to add the radiative tail following Braun et al.}
\hspace*{\parindent}
This model follows the regularization of Braun et al. as illustrated by equation (\ref{110e}), but using the full radiative tail of Lee et al. (\ref{115e}). We set
\beq
\label{136e}
\varphi_+^B (\omega ) = N \varphi_+^{BT} (\omega ) + \varphi_+^{RAD} (\omega )
\eeq

\noi where $\varphi_+^{BT} (\omega )$ is the same function (\ref{31e}) or (\ref{34e}) as in (\ref{116e}) and we now regularize the radiative tail by making the replacement ${1 \over \omega} \to {\omega \over \omega^2 + \mu^2}$, i.e. we take
\bea
\label{137e}
&&\varphi_+^{RAD} (\omega ) = {C_F \alpha_s \over  \pi} \ {\omega \over \omega^2 + \mu^2} \left \{ \left [ {1 \over 2} - \ {\rm log} \left ({\omega \over \mu}\right )  \right ]\right .  \nn \\
&&\left . + \  {4  \overline{\Lambda}_{SF}(\mu ) \over 3} \ {\omega \over \omega^2 + \mu^2}\left [ 2 - \ {\rm log} \left ({\omega \over \mu}\right ) - 6 \right ] \right \}
\eea

\noi The radiative tails (\ref{117e}) and (\ref{137e}) differ at low $\omega$. Now we do not have to determine the gluing point $\omega_t$. To impose the OPE conditions we proceed as follows. We have two conditions to fulfill 
\beq
\label{138e}
M_0^{model}(\Lambda_{UV}) = M_0^{OPE}(\Lambda_{UV})\qquad\qquad M_1^{model}(\Lambda_{UV}) = M_1^{OPE}(\Lambda_{UV})
\eeq

\noi These conditions give, respectively
\bea
\label{139e}
&&N = 1 + {C_F \alpha_s \over 4 \pi} \left ( \left [ - 2\  {\rm Log}^2 \left ( {\Lambda_{UV} \over \mu}\right ) + 2\  {\rm Log} \left ({\Lambda_{UV} \over \mu}\right ) - {\pi^2 \over 12}\right ]\right . \nn \\
&&+ \ {16  \overline{\Lambda}_{SF} \over 3\Lambda_{UV} }  \left [ {\rm Log} \left ({\Lambda_{UV} \over \mu}\right ) - 1 \right ] \\
&&\left . -\int_0^{\Lambda_{UV}} {4 \omega \over \omega^2 + \mu^2} \left \{ \left [ {1 \over 2} - \ {\rm log} \left ({\omega \over \mu}\right )  \right ] +  {4  \overline{\Lambda}_{SF}\over 3} \ {\omega \over \omega^2 + \mu^2}\left [ 2 - \ {\rm log} \left ({\omega \over \mu}\right )  \right ] \right \}\right )\nn 
\eea

$$M_+^{(1)BT} =  - \int_0^{\Lambda_{UV}} {C_F \alpha_s \over 4 \pi} \ {4 \omega \over \omega^2 + \mu^2} \left \{ \left [ {1 \over 2} - \ {\rm log} \left ({\omega \over \mu}\right )  \right ] +  {4  \overline{\Lambda}_{SF}\over 3} \ {\omega \over \omega^2 + \mu^2}\left [ 2 - \ {\rm log} \left ({\omega \over \mu}\right )  \right ] \right .$$
\bea
\label{140e}
&&+ \ \Lambda_{UV} \ {C_F \alpha_s \over 4 \pi} \left [ - 4\ {\rm Log} \left ({\Lambda_{UV} \over \mu}\right ) + 6 \right ] \\
&&+  \ {4  \overline{\Lambda}_{SF}\over 3}  \left ( 1 + {C_F \alpha_s \over 4 \pi} \left \{  -6\  {\rm Log} \left ({\Lambda_{UV} \over \mu}\right ) - {16  \overline{\Lambda}_{SF}\over 3 \Lambda_{UV} }  \left [ {\rm Log} \left ({\Lambda_{UV} \over \mu}\right ) -1 \right ] \right \} \right . \nn \\
&&\left . + \ {C_F \alpha_s \over 4 \pi}  \int_0^{\Lambda_{UV}} {4 \omega \over \omega^2 + \mu^2}  \left \{ \left [  {1 \over 2} - \  {\rm log} \left ( {\omega \over \mu}\right )  \right ] +  {4  \overline{\Lambda}_{SF}\over 3} \ {\omega \over \omega^2 + \mu^2}\left [ 2 - \ {\rm log} \left ({\omega \over \mu}\right )  \right ] \right \} \right ) \nn \\
&&-\  {C_F \alpha_s \over 4 \pi}\ 6 \mu\nn
\eea

\noi where $M_+^{(1)BT} = 0.964$~GeV from (\ref{108e}).\par

We have to solve these two equations for $N$ and $\overline{\Lambda}_{SF}$ in terms of $\Lambda_{UV}$ and find in which region of $\Lambda_{UV}$ these quantities become approximately constant. We find that this is the case for the reasonable range $\Lambda_{UV} > 3$~GeV, giving
\beq
\label{141e}
N = 0.856 \qquad\qquad  \overline{\Lambda}_{SF}(\mu ) = 0.563\ {\rm GeV} \qquad\qquad  \overline{\Lambda} = 0.338~{\rm GeV}
\eeq

\noi i.e.  a situation very close to the first model for $\overline{\Lambda}_{SF}(\mu )$. \par

Although the radiative tails are different, the final functions are almost identical, as shown in Fig. 4. The reason is that the positive part of the radiative tail in the latter model is compensated by the lower value of $N$ (\ref{141e}), imposed by the OPE constraints.\par

In this second model we have, for the quantities of interest, 
\beq
\label{142e}
\lambda_B^{-1} = 1.432\ {\rm GeV}^{-1} \qquad\qquad \sigma_B = 1.219
\eeq

\noi These values are very close to the ones of the first model (\ref{134e}), (\ref{135e}). \par

We conclude that our results, due to the OPE constraints, are invariant relatively to the way of adding the radiative tail to the non-perturbative part.

\section{Proposals for LCDA in other approaches.}
\hspace*{\parindent}
The alternative theoretical method for the calculation of the LCDA is essentially the one of QCD Sum Rules. Here we will distinguish between the work that considers the LCDA at leading order and the one incorporating the radiative tail. Our aim is not a critical one, but only to show the great variety of ans\"atze that one can find in the literature for the functions $\varphi_{\pm}^B(\omega )$, and the corresponding varied results for the parameters $\overline{\Lambda}$ and $\lambda_B^{-1}$.

\subsection{LCDA at leading order.}
\hspace*{\parindent}
A word of caution is in order here. In this part, that involves the LCDA at leading order, without radiative corrections, $\overline{\Lambda}$ is considered to be related to the center-of-gravity of the function $\varphi_{+}^B(\omega )$, i.e. to its first moment through
\beq
\label{143e}
<\omega >_+ \ = {4 \overline{\Lambda} \over 3}
\eeq

\noi Actually, in the calculation of the present paper, this first moment is assimilated to ${4 \overline{\Lambda}_{SF} \over 3}$, since the parameters are obtained from the fit of the Roman shape function to $B \to X_s \gamma$, that provides $\overline{\Lambda}_{SF}$. This will make easier the comparison with the other approaches in the absence of radiative corrections.\par

Before going to specific theoretical schemes, it is worth to quote the bound found by Korchemsky, Pirjol and Yan \cite{5r}, that is independent of the precise form of $\varphi_{+}^B(\omega )$, and follows from assuming positivity $\varphi_{+}^B(\omega ) > 0$
\beq
\label{144e}
\lambda_B^{-1} \geq {3 \over 4 \overline{\Lambda}}
\eeq

\noi Of course, the positivity condition $\varphi_{+}^B(\omega ) > 0$ is violated by the radiative tail, as we have seen in the preceding section.

\subsubsection{QCD Sum Rules.}
\hspace*{\parindent}
Grozin and Neubert did obtain from the QCDSR result the simple form for the functions $\varphi_{\pm}^B(\omega )$ \cite{2r}, 
\beq
\label{145e}
\varphi_{+}^B(\omega ) = {\omega \over \omega_0^2}\ e^{-\omega /\omega_0} \qquad \varphi_{-}^B(\omega ) = {1 \over \omega_0^2}\ e^{-\omega /\omega_0}
\eeq

\noi that satisfy relation (\ref{39e}) and give, for the positive moments (see also the analysis of ref. \cite{31r}), 
\beq
\label{146e}
<\omega >_+ \ =  2 <\omega >_- \ = 2 \omega_0 =  {4 \overline{\Lambda} \over 3}
\eeq

\noi and for the parameter $\lambda_B^{-1}$
\beq
\label{147e}
\lambda_B^{-1} = {1 \over \omega_0} = {3 \over 2 \overline{\Lambda}}
\eeq

From the value $\overline{\Lambda} \cong 0.55$~GeV used in this paper one gets
\beq
\label{148e}
\lambda_B^{-1} \cong 2.72\ {\rm GeV}^{-1}
\eeq

On the other hand, Braun et al. \cite{7r} have obtained from QCDSR the following simple form for the normalized long distance shape of the LCDA $\varphi_{+}^B(\omega )$
\beq
\label{149e}
\varphi_{+}(\omega ) = {3 \over 4 \varepsilon_c^3}\ \omega \left ( 2  \varepsilon_c - \omega \right ) \theta \left ( 2 \varepsilon_c - \omega \right )
\eeq
\vskip 5 truemm

\noi where $\varepsilon_c$ is the continuum threshold, and obtain
\beq
\label{150e}
<\omega >_+ \ =  \ 2<\omega >_- \ = {4 \overline{\Lambda} \over 3}\qquad\qquad  \lambda_B^{-1} = {3 \over 2\varepsilon_c} = {9 \over 8 \overline{\Lambda}}
\eeq

\noi that gives, for $\varepsilon_c = (0.9 \pm 0.1)$~GeV \cite{7r}
\beq
\label{151e}
\overline{\Lambda} \cong (0.67 \pm 0.07) \ {\rm GeV} \qquad\qquad  \lambda_B^{-1} \cong (1.68 \pm 0.18) \ {\rm GeV}^{-1}
\eeq

Also, within the QCDSR approach, Ball and Kou \cite{8r} find the above relation and the numerical value, adopting the value $ \overline{\Lambda} = 0.68$~GeV \cite{35r,36r}
\beq
\label{152e}
\lambda_B^{-1} = {9 \over 8  \overline{\Lambda}} \cong 1.67\ {\rm GeV}^{-1}
\eeq

Both refs. \cite{7r,8r} find the same expression (\ref{149e}). However, we must emphasize that in the work of Braun et al. \cite{7r}, formula (\ref{149e}) represents just the simplest contribution to the QCD Sum Rules. The full result of \cite{7r} is much more complicated, and the formula (\ref{110e}) represents a simple parametrization of the full expression. In ref. \cite{7r} the long distance part and the radiative tail follow together from the calculation, and the latter is not added by hand.

\subsubsection{QCD Factorization models.}
\hspace*{\parindent}
Within the QCD factorization approach of Beneke et al. \cite{3r} for the calculation of charmless non-leptonic $B$ decays ($\overline{B} \to \overline{K}\pi$, $\overline{B} \to \pi\pi , \cdots$) the following range is adopted~:
\beq
\label{156e}
\lambda_B^{-1} = (3.5 \pm 1.5)\  {\rm GeV}^{-1}
\eeq

\noi This is a guess essentially based on the determination (\ref{147e}) of \cite{2r}, using $\overline{\Lambda} \cong 0.4$~GeV. The same range is adopted in \cite{6r} for the description of the decay $B^- \to \gamma \ell \overline{\nu}_{\ell}$.

\subsubsection{pQCD Factorization models.}
\hspace*{\parindent}
A number of models have been proposed for the function $\Phi_+^B(\xi )$ defined in (\ref{3e}) within the framework of the Perturbative QCD Factorization (pQCD Factorization) approach, needed in the description of non-leptonic charmless $B$ decays. Integrating over ${\bf p}_{\bot}$ or equivalently taking impact parameter ${\bf b} = 0$, the following models were proposed \cite{36r,37r,38r} for the function $\Phi_+^{B}(\xi )$ defined by (\ref{3e}), 
\beq
\label{157e}
\Phi_+^{B}(\xi ) = N\ F(\xi ) \exp \left [ - {1 \over 2} \left ( {m_B \xi \over \omega_B}\right )^2 \right ]
\eeq

\noi with a range of values for $\omega_B$ in the interval
\beq
\label{158e}
0.25 \leq \omega_B \leq 0.65
\eeq

\noi and different models for the function $F(\xi )$,
\beq
\label{159e}
F( \xi ) = \xi^2 (1 - \xi )^2, \ \xi (1 - \xi ) \ \hbox{or} \ \sqrt{\xi (1 - \xi )}
\eeq

\noi with $N$ determined by the normalization condition (\ref{4e}). \par

To compare with the present work we have to make the change of variables (\ref{5e}) and take the limit $m_b \to \infty$ at $\omega$ fixed. Using relation (\ref{7e}) one finds the following models for $\varphi_+^B (\omega )$,
\beq
\label{160e}
\varphi_+^B (\omega ) = N\ G(\omega ) \exp \left [ - {1 \over 2} \left ( {\omega \over \omega_B}\right )^2 \right ]
\eeq

\noi with
\beq
\label{161e}
G(\omega) = \omega^2  ,\ \omega \ \hbox{or} \ \sqrt{\omega }
\eeq

\noi and $N$ determined by the normalization condition (\ref{8e}).\par

Just a few comments are in order to compare with the present work. First, in the BT scheme with an harmonic oscillator potential, $\varphi_+^B (\omega )$ is not proportional to a gaussian, but given by the function (\ref{107e}). Second, the models (\ref{157e})-(\ref{161e}) give generically low values for the first positive moment and very large values for the first inverse moment
\bea
\label{162e}
&&0.38\ {\rm GeV} \leq \ <\omega >_+ \ =  {4  \overline{\Lambda}\over 3} \leq 0.48\ {\rm GeV} \nn \\
&&2.66 \ {\rm GeV}^{-1}  \leq \ <\omega^{-1} >_+ \ =  \lambda_B^{-1} \leq 4.18
\eea

\subsubsection{Results in the BT quark model.}
\hspace*{\parindent}
The Bakamjian-Thomas quark model of the present paper with harmonic oscillator wave function and parameters taken from the fit to the $B \to X_s \gamma$ spectrum using the Roman Shape Function yields 
\beq
\label{163e}
<\omega >_+ \ = 2 <\omega >_- \ = {4  \overline{\Lambda}_{SF} \over 3} = 0.964\ {\rm GeV}  \eeq
\noi and
\beq
\lambda_B^{-1}  = \ <\omega^{-1}>_+\  = 1.521\ {\rm GeV}^{-1}
\label{164e}
\eeq

\noi that satisfies the bound (\ref{143e}). The comparison between $\lambda_B^{-1}$ and $<\omega >_+^{-1}$ is different than in the Lee et al. model (\ref{145e}), since we find approximately
\beq
\label{165e}
 <\omega^{-1}>_+ \ \cong {3 \over 4}   <\omega >^{-1}_+
\eeq

\noi $\lambda_B^{-1}$ is smaller than in the Lee-Neubert model due to the effect of the dynamical light quark mass $m_2$, that has a large value from the fit with the Roman Shape Function, and depresses $<\omega^{-1}>_+$ due to the behaviour (\ref{42e}).\par

From the value for $\overline{\Lambda}_{SF} = 0.723$~GeV from (\ref{163e}), not inconsistent with (\ref{120e}) we obtain, for $\mu = 1.5$~GeV, from (\ref{119e}) and (\ref{118e})~:
\beq
\label{NEW166e}
\overline{\Lambda} = 0.49\ {\rm GeV}
\eeq

We summarize the values obtained for the parameters $\overline{\Lambda}$ and $\lambda_B^{-1}$, guessed or used in other approaches, that we compare with the results of the present paper (Fig. 1), in Table 1 and in Fig.~5.\par
\begin{figure}
\centering\leavevmode
\includegraphics[scale=1.4]{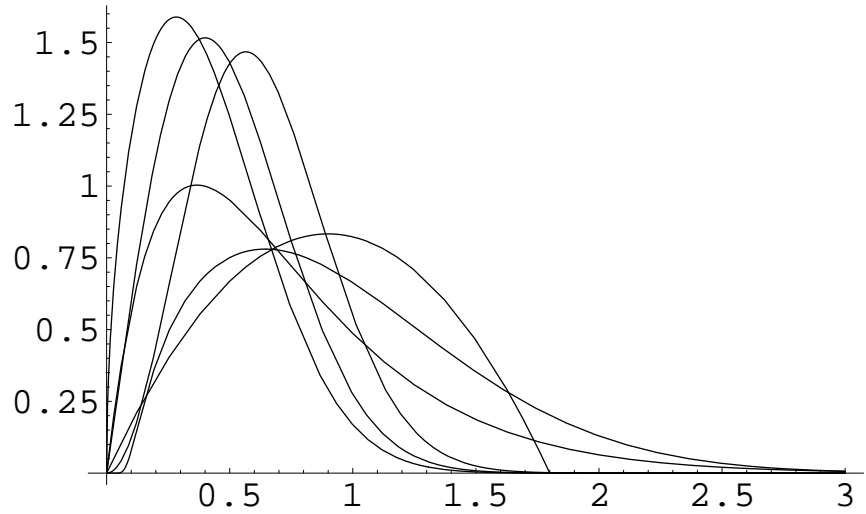}

\caption{Different models for the function $\varphi_+^B(\omega )$ in leading order. From higher to lower curves~: the heavy quark limit of the three models of pQCD \protect{\cite{36r, 37r,38r}} (\ref{160e}), (\ref{161e}) with $\omega_B = 0.4$~; follows the models of QCDSR \protect{\cite{2r}} (\ref{145e}) with $\overline{\Lambda} = 0.55$~GeV and \protect{\cite{7r,8r}} (\ref{149e})~; the wider curve is the model of the present paper (\ref{31e}) with the harmonic oscillator wave function (\ref{86e}) and the parameters (\ref{99e}), that give $\overline{\Lambda} = 0.49$~GeV. 
\centerline{\hbox to 8 truecm{\hrulefill}}}
\label{fig.5}
\end{figure}

A remark is in order here in the comparison in Fig.~5 between $\varphi_+^B(\omega )$ from, e.g. the QCDSR result (\ref{145e}) and our model. While in the former $<\omega>_+\ = {4 \overline{\Lambda} \over 3}$ with $\overline{\Lambda} = 0.55$~GeV, we have in our model (\ref{163e}), $<\omega>_+\ = {4 \overline{\Lambda}_{SF} \over 3}$ with $\overline{\Lambda}_{SF} = 0.723$~GeV, since it is $\overline{\Lambda}_{SF}$ that is determined by the Shape Function. This explains why our $\varphi_+^B(\omega )$ is more spread than (\ref{145e}).

\begin{center}
\begin{tabular}{|c|c|c|}
\hline
 Theoretical frame  &$\overline{\Lambda}$ &$\lambda_B^{-1}$ \\
\hline
Positivity & &$\geq {3  \over 4 \overline{\Lambda}}$ \\
$\varphi_+^B (\omega ) \geq 0$ \protect{\cite{5r}}& &\\
\hline
QCDSR \protect{\cite{2r}} &0.55 GeV &${3  \over 2 \overline{\Lambda}} = 2.72$ GeV$^{-1}$ \\
\hline
QCD Factorization &$\sim$ 0.45 GeV &${3  \over 2 \overline{\Lambda}} = (3.5 \pm 1.5)$ GeV$^{-1}$\\
\protect{\cite{3r,6r}} & & \\ 
\hline
QCDSR \protect{\cite{7r,8r}} &$(0.67 \pm 0.07)$ GeV &${9  \over 8 \overline{\Lambda}} = (1.7 \pm 0.2)$ GeV$^{-1}$ \\
\hline
pQCD \protect{\cite{36r}}-\protect{\cite{38r}} &$(0.32 \pm 0.04)$ GeV &$(3.42 \pm 0.76)$ GeV$^{-1}$ \\
\hline
BT model with & & \\
Roman shape &0.49 GeV &1.52 GeV$^{-1}$\\
function parameters & & \\
\hline
\end{tabular}
\end{center}

\noi Table 1 : Results for the parameters $\overline{\Lambda}$ and $\lambda_B^{-1}$ in the different theoretical approaches for $\varphi_+^B (\omega )$ in the absence of the radiative tail. 
\subsection{LCDA with a radiative tail.}
\hspace*{\parindent}
Let us now discuss the impact of adding a radiative tail to the LCDA in the various approaches. Braun et al \cite{7r} have proposed the parametrization (\ref{110e}) of $\varphi_+^B (\omega , \mu )$ for $\mu = 1$ including the radiative tail. Lee and Neubert \cite{9r} take as a model the long distance piece expression (\ref{145e}) adding the tail (\ref{117e}) with $\overline{\Lambda}_{SF}(\mu )$ replaced by $\overline{\Lambda}_{DA}(\mu , \mu )$,
\beq
\label{167e}
\varphi_{+}^B (\omega ) = N {\omega \over \omega_0^2}  \ e^{-\omega /\omega_0} + \theta \left ( \omega - \omega_t\right )  \varphi_+^{RAD} (\omega )
\eeq

The consistency of (\ref{167e}) with the first moments with the OPE (\ref{112e})-(\ref{113e}) imposes constraints on $N$ and $\omega_0$ that only depend on $\omega_t$ and $\overline{\Lambda}_{DA} = \overline{\Lambda}_{DA}(\mu , \mu )$ \cite{9r}. For $\mu = 1$, using the relations between $\overline{\Lambda}_{DA}(\mu , \mu )$, $\overline{\Lambda}_{SF}(\mu_* , \mu_* )$ and $\overline{\Lambda}$ one finds from $\overline{\Lambda}_{SF}(\mu_* , \mu_* ) = 0.65$~GeV (\ref{120e}), $\overline{\Lambda}_{DA}(\mu , \mu ) = 0.52$~{\rm GeV} $(\mu = 1$~GeV), $\omega_t = 2.33$~GeV, $N = 0.963$, $\omega_0 = 0.438$ and the results of Table~5. On the other hand, Lee et al. have shown that the model of Braun et al. for $\varphi_+^B (\omega )$, for $\mu = 1$~GeV is quite close to their own.\par

We show in Fig. 6 and Table 2 the results of our models for adding the radiative tail of sections 9.1 and 9.2, compared to the results of \cite{7r} and \cite{9r}. 

\begin{center}
\begin{tabular}{|c|c|c|}
\hline
 Method &$\overline{\Lambda}$ (GeV)  &$\lambda_B^{-1}$ (GeV$^{-1}$) \\
\hline
Braun et al. \protect{\cite{7r}} &0.4 - 0.5 &$1.98 \pm 0.52$  \\
\hline
Lee-Neubert \protect{\cite{9r}} &0.34 &$1.86 \pm 0.17$  \\
\hline
BT model (\ref{1e}) &0.35 &1.43  \\ 
\hline
BT model (\ref{2e}) &0.34 &1.43  \\ 
\hline
\end{tabular}
\end{center}

\noi Table 2 : Results for the parameters $\overline{\Lambda}$ and  $\lambda_B^{-1}$ for $\mu = 1.5$~GeV in the QCDSR approach and in our models (\ref{1e}) and (\ref{2e}) of sections 9.1 and 9.2 including the radiative tail.
\par \vskip 5 truemm

Some comments are in order here on the row of ref. \cite{7r} in Table 5. The range for $\overline{\Lambda}$  is the choice given in \cite{7r}, below formula (\ref{15e}). The value of $\lambda_B^{-1}$ (1.5 GeV) is obtained from $\lambda_B^{-1}$ (1. GeV) (formula (39) of \cite{7r}) using $\varphi_+ (\omega , \mu = 1 \ {\rm GeV})$ (formulas (\ref{39e}), (\ref{43NEWEQ}) and (\ref{43e})) and the scale dependence for $\lambda_B^{-1}(\mu )$ given by (\ref{41e}).

\begin{figure}
\centering\leavevmode
\includegraphics[scale=1.4]{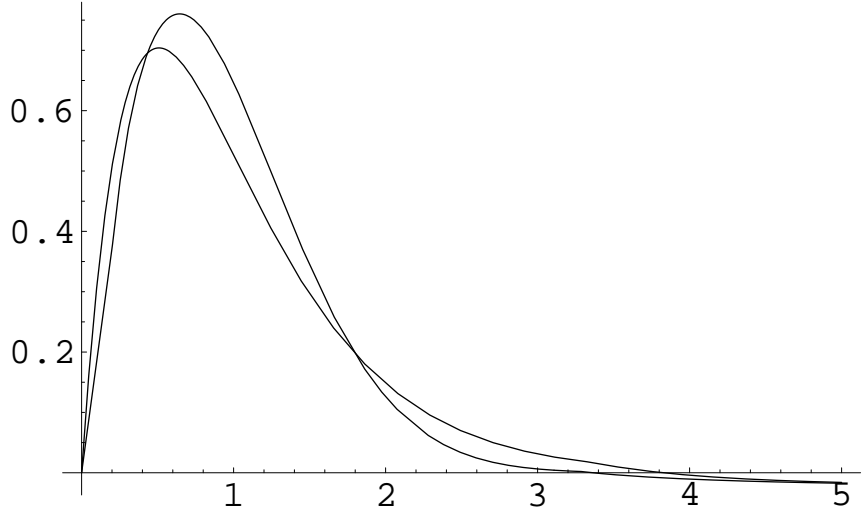}
\caption{The function $\varphi_+^B(\omega )$ including the radiative tail for $\mu = 1.5$~GeV. From higher to lower curves~:  model of the present paper~; Lee-Neubert model \protect{\cite{9r}}.
\centerline{\hbox to 8 truecm{\hrulefill}}}
\label{fig.6}
\end{figure}

\section{Phenomena sensitive to the LCDA.}
\hspace*{\parindent}
Let us now review the observables that are related to the LCDA. In this respect there is essentially, the work based on QCD Factorization and the one based on QCD Sum Rules. We will briefly review~:\par

(1) The decay $B^- \to \gamma \ell \overline{\nu}_{\ell}$.\par

(2) Hard scattering contribution to non-leptonic decay amplitudes like  $A(\overline{B}_d \to \pi^+\pi^- )$ in the framework of QCD Factorization.\par

(3) Asymptotic behaviour of the Isgur-Wise function $\xi (w)$ and the subleading form factor $\xi_3 (w)$.

(4) Heavy-to-light form factors like $B \to \pi$ at $q^2 = 0$.

\subsection{The decay B$^{\bf -} \to \gamma \ell \overline{\nu}_{\ell}$.}
\hspace*{\parindent}
This decay is described by two form factors $F_V (E_{\gamma})$ and $F_A(E_{\gamma})$ that, at tree level, are related, in the heavy quark limit, to the LCDA $\varphi_+^B (\omega )$ by
\beq
\label{185e}
F_V (E_{\gamma}) = F_A (E_{\gamma}) = {f_B m_B Q_u \over 2 E_{\gamma}} \ <\omega^{-1}>_+ \ = {f_B m_B Q_u \over 2 E_{\gamma}} \ {1 \over \lambda_B}
\eeq

\noi This process is directly related to the parameter $\lambda_B^{-1}$ and would be the most direct way of measuring it. Conversely, having a good theoretical estimate of $\lambda_B^{-1}$, the process $B^- \to \gamma \ell \overline{\nu}_{\ell}$ allows to measure $|V_{ub}|$. \par

A considerable effort has been devoted to the study of this decay going beyond the tree result (\ref{185e}). Korchemsky et al. \cite{5r} have computed the form factors for photon energies larger than $\Lambda_{QCD}$ combining QCD methods for exclusive processes with HQET. They have written the leading twist form factors as the convolution of the $B$ meson light-cone amplitude $\varphi_+^B(\omega )$ with a hard scattering term, and computed also Sudakov contributions. In a later paper, Braun et al. \cite{7r} have considered the radiative tail for the LCDA $\varphi_+^B (\omega )$, reviewed in Sections 9 and 10. Descotes-Genon and Sachrajda \cite{6r} have studied the decay $B^- \to \gamma \ell \overline{\nu}_{\ell}$ in the framework of QCD factorization, demonstrating that indeed at the one loop order the amplitude can be written as a convolution of a perturbatively calculable hard-scattering amplitude with $\varphi_+^B(\omega )$. For the parameter $\lambda_B$, they use the guess of \cite{3r} $\lambda_B = (350 \pm 150)$~MeV. \par

The scheme of the present paper predicts a value $\lambda_B^{-1} \cong 1.43$~GeV$^{-1}$ that is in the lower range given in the different schemes of the literature, as reviewed in Section 10. This feature is due to the rather large value of the dynamical mass of the constituent light quark. \par

It is important to underline that the decay $B^- \to \gamma \ell \overline{\nu}_{\ell}$ is the only process that could allow, in principle, to directly measure $\lambda_B^{-1}$, modulo radiative and $1/m_Q$ corrections.\par

Recently, in a search for the decay $B^+ \to \gamma \ell^+ \gamma_{\ell}$, with $\ell = e$ or $\mu$ BaBar has found the following upper bounds \cite{39newr}, depending on the way of analyzing the data~:
\beq
\label{Neweq1}
\lambda_B^{-1} < 1.49\ {\rm GeV}^{-1}
\eeq

\noi or
\beq
\label{Neweq2}
\lambda_B^{-1} < 1.69\ {\rm GeV}^{-1}
\eeq

\noi These bounds are fulfilled by our value (\ref{134e}) $\lambda_B^{-1} = 1.43\ {\rm GeV}^{-1}$ while they seem at odds with the predictions of other schemes (Tables 4 and 5). The predictions of refs. \cite{7r} and \cite{9r} are within $1 \sigma$ or $2 \sigma$ in agreement with the bounds (\ref{Neweq1}) or (\ref{Neweq2}).

\subsection{Hard scattering in non-leptonic two-body B decays.}
\hspace*{\parindent}
The correction to factorization to the decays with an emitted light meson $\overline{B} \to D\pi, \pi \pi$, ... , that comes from a gluon attached to the spectator quark, called the Hard Scattering Amplitude, depends directly on $\Phi_+^B (\xi )$ or $\varphi_+^B (\omega )$ \cite{3r}. In the case of the decays to two light mesons, it scales in terms of dimensionfull quantities like $G_F\alpha_s m_b \Lambda_{QCD}^{5/2}$, having the same behaviour as the leading term $G_F m_b \Lambda_{QCD}^{5/2}$. In this case of decays of $B$ to two light mesons, this results in a contribution to the effective QCD factors like $a_1$ \cite{3r}
\beq
\label{174e}
a_{1, \parallel} = {C_2 \over N_c}\ {C_F \pi \alpha_s \over N_c} \ H_{K\pi}
\eeq

\noi with
\beq
\label{175e}
H_{K\pi} = {1 \over \lambda_B}\ {f_B f_{\pi} \over m_B F_0^{B \to \pi} (0)} \left [ <(1-x)^{-1}>_K\ <(1-y)^{-1}>_{\pi}\ + \ r_{\chi}^{\pi}  <x^{-1}>_K X_H^{\pi}\right ]
\eeq

\noi where $r_{\chi}^{\pi}$ is the well-known chiral enhancement factor \cite{3r}. This contribution to non-leptonic decays is proportional to $\lambda_B^{-1}$. BBNS \cite{3r} propose the number $\lambda_B = (350 \pm 150)$~MeV, based on the relation to $\overline{\Lambda}$ obtained in \cite{2r} and the bound \cite{5r}. The range adopted by BBNS $\lambda_B^{-1} = (3.5 \pm 1.5)$~GeV$^{-1}$ is larger than the prediction of the model of the present paper, $\lambda_B^{-1} \cong 1.43$ or $\lambda_B^{-1} \cong 1.52$ with or without the radiative tail. However, in the phenomenological analysis of two-body non-leptonic decays, as we can see in expression (\ref{175e}), $\lambda_B^{-1}$ is afffected by a subleading although important term, chirally enhanced and proportional to the unknown logarithmically divergent factor $\int_0^1 {dy \over 1 - y} \Phi_p^{\pi} (y)$, where $ \Phi_p^{\pi} (y) \cong 1$ is the twist-3 pion light-cone amplitude. The second term in (\ref{175e}) is parametrized phenomenologically by $X_H^{\pi} = (1 + \rho_H e^{i\varphi_H} ){\rm Log} (m_B/\mu )$ ($\rho_H \leq 1$), and fitted to the data. The modulus of the ratio between the two terms in (\ref{175e}) is roughly $|r_{\chi}^{\pi} X_H^{\pi}| = {2 m_{\pi}^2 \over m_b (m_u + m_d)} |1 + \rho_H e^{i\varphi_H}| {\rm Log} (m_B/\mu )$, of $O(1)$. This term results in a large uncertainty on $H_{K\pi}$ (Fig.~5 of \cite{3r}), that for the real part is $0.5 \leq {\rm Re} H_{K\pi} \leq 2.5$, with a somewhat smaller uncertainty for the imaginary part. Therefore, strictly speaking, due to this unknown term, that plays a non-negligible role in the description of the data, two-body non-leptonic decays do not allow to make a model-independent extraction of the parameter $\lambda_B^{-1}$. We can conclude that data on non-leptonic decays into two light mesons, although dependent on $\lambda_B^{-1}$, are not a model independent determination of this quantity.

\subsection{Asymptotic behaviour of the functions $\xi(w)$ and $\xi_3(w)$.}
\hspace*{\parindent}
Grozin and Neubert \cite{2r} have given the behaviour of the elastic Isgur Wise function $\xi (\cosh \theta)$ in the large recoil limit $w \gg 1$ ($\theta \gg 0$),
\beq
\label{176e}
\xi (\cosh \theta) = 16 \pi \alpha_s {C_F \over N_c} \ f^2 \ <\omega^{-2}>_+\ <\omega^{-1}>_-\ e^{-2 \theta}
\eeq

\noi where $2f = f_M \sqrt{m_Q}$. In our notation, this writes
\beq
\label{177e}
\xi (w) \cong \pi \ {C_F \alpha_s \over N_c} \left ( f_B \sqrt{m_B}\right )^2 \ <\omega^{-2}>_+\ <\omega^{-1}>_-\ {1 \over w^2}
\eeq

\noi where
\beq
\label{178e}
<\omega^{-2}>_+\ = \int_0^{\infty} {d \omega \over \omega^2}\ \varphi_+^B (\omega ) \qquad\qquad <\omega^{-1}>_-\  = \int_0^{\infty} {d \omega \over \omega}\ \varphi_-^B (\omega ) 
\eeq

\noi From the description of the behaviour of the functions $\varphi_\pm^B (\omega )$ in the different limits in Sections 6 and 10, one realizes that these two integrals diverge for massless light quarks. However, the non-vanishing light quark mass of the BT model, implying the behaviour (\ref{42e}), provides a natural infrared cut-off. \par

Using expressions (\ref{31e}) or (\ref{34e}) for non-vanishing light quark mass and the wave function (\ref{86e}) with the parameters (\ref{99e}), one finds the finite results
\beq
\label{179e}
<\omega^{-2}>_+ \ = 3.826\ {\rm GeV}^{-2} \qquad\qquad  <\omega^{-1}>_- \ = 2.322\ {\rm GeV}^{-1}
\eeq

\noi From the value $f_B \sqrt{m_B}= 0.388$~{\rm GeV}$^{3/2}$ found in Section 8.2 in the present case, one finds the behaviour
\beq
\label{180e}
\xi (w) \cong 1.337 \ \pi \ {C_F \alpha_s \over N_c}\ {1 \over w^2}
\eeq
\vskip 5 truemm

\noi This gives, for $\mu = 1$ ($C_F \alpha_s \cong 0.624$), an order of magnitude $\xi (w) \sim {1 \over w^2}$, smaller but of the same order of magnitude as with a function $\xi (w) \cong \left ({2 \over w + 1}\right )^2$ and a slope of the order $\rho^2 \cong 1$.\par

A comment is in order here. Grozin and Neubert \cite{2r} argue that the logarithmic singularities of  $<\omega^{-2}>_+$ and $<\omega^{-1}>_-$ would be cut off by the transverse momenta and virtualities of the light quarks in the mesons. In our scheme, the transverse momenta do not play such a role, since the divergence remains taking those into account, as shown by the explicit expressions for $\varphi_\pm^B (\omega )$ in the massless limit that can be read from (\ref{31e}) or (\ref{34e}). It is the non-vanishing dynamical light quark mass, that one can consider as a ``virtuality'' of the massless quark due to the $<\overline{q}q>$ condensate, that ensures the finiteness of the moments $<\omega^{-2}>_+$ and $ <\omega^{-1}>_-$. \par

The subleading function $\xi_3(w)$ \cite{39r}, coming from $1/m_Q$ perturbations to the current, behaves \cite{2r} in the large recoil limit $\theta \gg 0$ like
\beq
\label{181e}
\xi_3 (\cosh \theta ) = 4 \pi \alpha_s \ {C_F \over N_c}\ f^2 \ <\omega^{-1}>_+^2\ e^{-\theta }
\eeq

\noi or, in our notation,
\beq
\label{182e}
\xi_3 (w) \cong {\pi \over 2}\  {C_F \alpha_s \over N_c} \left ( f_B \sqrt{m_B}\right )^2 \ <\omega^{-1}>_+^2\ {1 \over w}
\eeq

\noi and from the value (\ref{134e}) or (\ref{142e}) for $<\omega^{-1}>_+ \ = \lambda_B^{-1}$,
\beq
\label{183e}
<\omega^{-1}>_+ \ \cong 1.521\ {\rm GeV}^{-1}
\eeq

\noi we find
\beq
\label{184e}
\xi_3 (w) \cong 0.174\ \pi\ {C_F \alpha_s \over N_c} \ {1 \over w}\ {\rm GeV}
\eeq

\noi that gives 
\beq
\label{NEW183EQ}
\xi_3(w) \sim {0.1 \over w} 
\eeq

\subsection{B $\to \pi$ form factor at q$^{\bf 2}$ = 0.}
\hspace*{\parindent}
From eq. (\ref{23e}) of \cite{1r} one has up to terms in $1/(1-u)$,
\beq
\label{168e}
F_{+,0}^{B \to \pi}(0) \cong {\pi \alpha_s C_F \over N_c}\ {f_{\pi} f_B \over m_b^2} \int_0^1 d\xi \ du \ \Phi_-^B (\xi ) \  \Phi_{\pi} (u) {1 \over \xi (1 - u)^2}
\eeq

\noi i.e.,
\beq
\label{169e}
F_{+,0}^{B \to \pi}(0) \cong {\pi \alpha_s C_F\over N_c}\ {f_{\pi} f_B \over m_b} \int_0^{\infty} {d\omega \over \omega} \varphi_- (\omega ) \int_0^1 {du \over (1-u)^2} \ \Phi_{\pi} (u) 
\eeq

\noi However, this expression is not usable due to the infrared divergence of the integral over $u$, since $\Phi_{\pi} (u) \cong 6u(1-u)$.\par

One way out of this problem is the use of the Light Cone QCD Sum Rules approach by Khodjamirian et al. \cite{4r}, that provides a simple and explicit expression for the heavy-to-light form factors in terms of LCDA. Taking as an example the form factor $F_{+}^{B \to \pi}(0)$ one has,
\beq
\label{170e}
F_{+}^{B \to \pi}(0) = {f_B \over  f_{\pi} m_B} \int_0^{s_0^{\pi }} ds \exp  \left ( - {s \over M^2}\right ) \varphi_- ^B \left (  {s \over m_B}\right )
\eeq

\noi An approximation to this expression has been used in the first reference of \cite{4r}, 
\beq
\label{171e}
 \int_0^{s_0^{\pi }} ds \exp  \left ( - {s \over M^2}\right ) \varphi_-^B \left ( {s \over m_B}\right ) \cong \varphi_- ^B(0)  \int_0^{s_0^{\pi }} ds \exp  \left ( - {s \over M^2}\right )
\eeq

\noi that gives, from the integral relation (\ref{40e}), valid only for {\it vanishing light quark mass},
\beq
\label{172e}
F_{+}^{B \to \pi}(0) = {1 \over \lambda_B}\ {f_B \over  f_{\pi} m_B} M^2 \left [ 1 - \exp  \left ( - {s_0^{\pi} \over M^2}\right ) \right ]
\eeq

\noi Some remarks are in order here~:\par

(1) If $\varphi_+^B (\omega )$ and $\varphi_- ^B(\omega )$ are related by (\ref{40e}), the numerical results for $F_{+}^{B \to \pi}(0)$ from (\ref{170e}) and (\ref{172e}) are very close. \par

(2) As pointed out in \cite{4r} and we observe from (\ref{172e}),  for a given window of the Borel parameter $M^2$ and a value of $s_0^{\pi}$, 
\beq
\label{173e}
0.5\ {\rm GeV}^2 \leq M^2 \leq 1.2\ {\rm GeV}^2\qquad\qquad s_0^{\pi} = 0.7\ {\rm GeV}^2 
\eeq
\noi $F_{+}^{B \to \pi}(0)$ is very sensitive to the precise value of $\lambda_B^{-1}$. The value adopted in \cite{4r} is  $\lambda_B^{-1} \cong 2$~GeV$^{-1}$.

(3) The differential relation (\ref{40e}) between $\varphi_+^B (\omega )$ and $\varphi_-^B (\omega )$ is only valid for $m_2 = 0$, and is badly violated in the BT model examined here, where the value of $m_2$ extracted from the $B \to X_s \gamma$ spectrum is rather large.\par

(4) A non-vanishing dynamical light quark mass $m_2$ has a dramatic impact since then $\varphi_-^B(0) = 0$. We are not allowed then to use  (\ref{172e}) but must rely on the relation (\ref{170e}). Using this expression and the parameters (\ref{173e}) one obtains a very small value for $F_{+}^{B \to \pi}(0) \sim 0.015$, the reason being that $\varphi_-^B(\omega )$ vanishes at $\omega = 0$. One would need a much larger value of $s_0^{\pi}$ to get an appreciable contribution of $\varphi_-^B(\omega )$ to the integral (\ref{170e}). Our conclusion is that with our prediction for $\varphi_-^B( \omega )$ (Fig.~2), relation (\ref{170e}) and the set of parameters (\ref{173e}), one cannot describe the form factor $F_+^{B\to \pi}$. This feature deserves further investigation.

\section{Conclusions.}
\hspace*{\parindent}
In conclusion, within the Bakamjian-Thomas relativistic quark model, that in the heavy quark limit yields covariant form factors and Isgur-Wise scaling, we have computed the $B$ meson Light Cone Distribution Amplitudes $\varphi_\pm^B (\omega )$, that are also covariant in this scheme, and satisfy, in the limit of vanishing dynamical light quark mass, the integral relation given by QCD in the valence quark-antiquark sector. We have also computed the Shape Function $S(\omega )$ that enters in the description of the decay $B \to X_s \gamma$. The Light Cone Distribution Amplitudes and the Shape Function are related in the BT class of models and given in terms of the $Q\overline{q}$ internal wave function. Using a gaussian wave function, we have shown that the shape function is identical to the so-called Roman Shape Function. Using the parameters of the latter that fit the $B \to X_s \gamma$ spectrum, we have predicted the LCDA $\varphi_\pm^B (\omega )$. We have discussed the role played by the dynamical mass of the light constituent quark and included the short distance behaviour of QCD for $\varphi_+^B (\omega )$. Compared to most schemes in the literature, our model predicts a rather small value for the parameter $\lambda_B^{-1} \cong 1.43$~GeV$^{-1}$, due to the rather large value of the constituent light quark mass, fitted from the $B \to X_s\gamma$ spectrum. This value for $\lambda_B^{-1}$ fulfills the upper bounds obtained by BaBar from the search of $B^+ \to \gamma \ell^+ \nu_{\ell}$. Moreover, the non-vanishing constituent light quark mass has the important implication $\varphi_-^B(0) = 0$. We have compared with other theoretical approaches and discussed the phenomena that are sensitive to the LCDA.

\section*{Acknowledgements.}
\hspace*{\parindent}
We have benefited from informative comments from S\'ebastien Descotes-Genon, from Emi Kou about the BaBar bounds on $\lambda_{B}^{-1}$, and from S. Simula on ref. \cite{13r}. This work was supported in part by the EU Contract No. MRTN-CT-2006-035482 (FLAVIANET).


\begin{thebibliography}{99}
\bibitem{1r} M. Beneke, G. Buchalla, M. Neubert and C.T. Sachrajda, hep-ph/0006124, Nucl. Phys. B {\bf 591}, 313 (2000).
\bibitem{2r} A. G. Grozin and M. Neubert, hep-ph/9607366, Phys. Rev. D {\bf 55}, 272 (1997).
\bibitem{3r} M. Beneke, G. Buchalla, M. Neubert and C.T. Sachrajda, hep-ph/0104110, Nucl. Phys. B {\bf 606}, 245 (2001).
\bibitem{4r} A. Khodjamirian, T. Mannel and N. Offen, hep-ph/0504091, Phys. Lett. B {\bf 620}, 52 (2005)~; hep-ph/0611193, Phys. Rev. D {\bf 75}, 054013 (2007).
\bibitem{5r} G. P. Korchemsky, D. Pirjol and T.-M. Yan, hep-ph/9911427, Phys. Rev. D {\bf 61}, 114510 (2000).
\bibitem{6r} S. Descotes-Genon and C. T. Sachrajda, hep-ph/0209216, Nucl. Phys. B {\bf 650}, 356 (2003).
\bibitem{7r} V. M. Braun, D. Y. Ivanov and G. P. Korchemsky, hep-ph/0309330, Phys. Rev. D {\bf 69}, 034014 (2004).
\bibitem{8r} P. Ball and E. Kou, hep-ph/0301135, JHEP {\bf 0304}, 029 (2003).
\bibitem{9r} S. J. Lee and M. Neubert, hep-ph/0509350, Phys. Rev. D {\bf 72}, 094028 (2005).
\bibitem{10r} B. Bakamjian and L. H. Thomas, Phys. Rev. {\bf 92}, 1300 (1953).
\bibitem{11r} B. D. Keister and W. N. Polyzou, Adv. Nucl. Phys. {\bf 20}, 225 (1991).
\bibitem{12r} M. Terent'ev, Sov. J. Nucl. Phys. {\bf 24}, 106 (1976).
\bibitem{13r} F. Cardarelli, I. L. Grach, I. M. Nadoretskii, E. Pace, G. Sale and S. Simula, Phys. Lett. B {\bf 332}, 1 (1994).
\bibitem{14r} A. Le Yaouanc, L. Oliver, O. P\`ene and J.-C. Raynal, Phys. Lett. B {\bf 365}, 319 (1996).
\bibitem{15r} N. Isgur, M. B. Wise, Phys. Rev. D {\bf 43}, 819 (1991).
\bibitem{16r} V. Mor\'enas, A. Le Yaouanc, L. Oliver, O. P\`ene and J.-C. Raynal, Phys. Rev. D {\bf 56}, 5668 (1997).
\bibitem{17r} A. Le Yaouanc, L. Oliver, O. P\`ene and J.-C. Raynal, Phys. Lett. B {\bf 386}, 304 (1996).
\bibitem{18r} A. Le Yaouanc, L. Oliver, O. P\`ene and J.-C. Raynal and V. Mor\'enas, Phys. Lett. B {\bf 520}, 25 (2001).
\bibitem{19r} A. Le Yaouanc, L. Oliver and J.-C. Raynal, hep-ph/0210233, Phys. Rev. D {\bf 67}, 114009 (2003)~; Phys. Lett. B {\bf 557}, 207 (2003).
\bibitem{20r} G. Altarelli et al., Nucl. Phys. B {\bf 208}, 365 (1982).
\bibitem{21r} I. Bigi, M. Shifman, N. Uraltsev and A. Vainshtein, hep-ph/9402225, Phys. Lett. B {\bf 328}, 431 (1994).
\bibitem{22r} T. Mannel and M. Neubert, hep-ph/9402288, Phys. Rev. D {\bf 50}, 2037 (1994).
\bibitem{23r} S. W. Bosch, B. O. Lange, M. Neubert and G. Paz, hep-ph/0402094, Nucl. Phys. B {\bf 699}, 335 (2004).
\bibitem{24r} S. W. Bosch, M. Neubert and G. Paz, hep-ph/0409115, JHEP {\bf 0411}, 073 (2004).
\bibitem{25r} B. O. Lange, M. Neubert and G. Paz, hep-ph/0504071, Phys. Rev. D {\bf 72}, 073006 (2005).
\bibitem{26r} A. Limosani and T. Nozaki (of the Belle Collaboration for the Heavy Flavor Averaging Group), hep-ex/0407052.
\bibitem{27r} S. Godfrey and N. Isgur, Phys. Rev. D {\bf 32}, 189 (1985).
\bibitem{28r} V. Mor\'enas, A. Le Yaouanc, L. Oliver, O. P\`ene and J.-C. Raynal, Phys. Lett. B {\bf 386}, 315 (1996).
\bibitem{29r} V. Mor\'enas, A. Le Yaouanc, L. Oliver, O. P\`ene and J.-C. Raynal, Phys. Rev. D {\bf 58}, 114019 (1998).
\bibitem{30r} M. Beneke and T. Feldman, Nucl. Phys. B {\bf 592}, 3 (2001).
\bibitem{31r} H. Kawamura, J. Kodaira, C. F. Qiao and K. Tanaka, hep-ph/0109181, Phys. Lett. B {\bf 523}, 111 (2001)~; Erratum - ibid. B {\bf 536}, 344 (2002).
\bibitem{32r} P. Koppenburg et al. (Belle Collaboration), hep-ex/0403004, Phys. Rev. Lett. {\bf 93}, 061803 (2004).
\bibitem{33r} M. Neubert, hep-ph/0412241, Phys. Lett. B {\bf 612}, 13 (2005)~; hep-ph/0506245, Phys. Rev. D {\bf 72}, 074025 (2005).
\bibitem{34r} A. H. Hoang, hep-ph/9905550, Phys. Rev. D {\bf 61}, 034005 (2000)~; A. Pineda, hep-ph/0105008, JHEP {\bf 0106}, 022 (2001).
\bibitem{35r} V. Lubicz, Nucl. Phys. Proc. Suppl. {\bf 94}, 116 (2001).
\bibitem{36r} Y.-Y. Keum, H. L. Li and A. I. Sanda, Phys. Lett. B {\bf 504}, 6 (2001).
\bibitem{37r} Y.-Y. Keum, H. L. Li and A. I. Sanda, Phys. Rev. D {\bf 63}, 054008 (2001).
\bibitem{38r} Y.-Y. Keum and H. L. Li, Phys. Rev. D {\bf 63}, 074006 (2001).
\bibitem{39newr} B. Aubert et al. (Babar Collaboration), arXiv:0704.1478 [hep-ex].
\bibitem{39r} A. F. Falk and M. Neubert, Phys. Rev. D {\bf 47}, 2965 (1993).
\end{thebibliography}
\end{document}